\documentstyle[12pt,epsf]{article}
\def\ehat{\hat{e}}

\def\al{\alpha}
\def\be{\beta}
 
\def\ep{\epsilon}
\def\lam{\lambda}

\def\calA{{\cal A}}   
\def\calD{{\cal D}}  
\def\calG{{\cal G}}  
\def\calJ{{\cal J}}  
  \def\calO{{\cal O}}

\def\calV{{\cal V}}  
 \def\calZ{{\cal Z}} 

\newsavebox{\prooffig}

\def\shead#1{\parbigskipn {\bf #1} \parmedskipn} 
\def\del        {  \partial  }
\def\half       {  {1\over 2}  }
\def\rootof#1   {  \left( #1 \right)^{1/2}  } 
\def\trace      {  \mbox{Tr}  }
\def\Tr         { {\rm Tr} }
\def\abs#1      {  \vert #1 \vert  }
\def\ie         {  {\it i.e.}      }
\def\evalat#1   {  \left\vert_{#1} \right. } 

\def\where      { \mbox{where}\qquad }
\def\comma          {\, ,}
\def\period         {\, .}
\def\lsim    {\lower .65ex \hbox{\ $\stackrel{<}{\sim}$\ } }
\def\gsim    {\lower .65ex \hbox{\ $\stackrel{>}{\sim}$\ } }
\def\vecii#1#2      {  \left(\begin{array}{c}#1\\#2\end{array}\right)  }
\def\veciii#1#2#3   {  \left(\begin{array}{c}#1\\#2\\#3\end{array}
                     \right)  }
\def\veciv#1#2#3#4  {  \left(\begin{array}{c}#1\\#2\\#3\\#4
                                 \end{array}\right)  }
\def\vecfv#1#2#3#4#5 {  \left(\begin{array}{c}#1\\#2\\#3\\#4\\#5
                                 \end{array}\right)  }

\def\matrixii#1#2#3#4            {  \left(\begin{array}{cc}#1&#2\\#3&#4
                                       \end{array}\right) }
\def\matrixiii#1#2#3#4#5#6#7#8#9 {  \left(\begin{array}{ccc}#1&#2&#3\\
                                     #4&#5&#6\\#7&#8&#9\end{array}
                               \right)  }
\def\mativ#1#2#3#4               {  \left(\begin{array}{cccc}
                                       #1\\#2\\#3\\#4\end{array}\right) }
\def\matv#1#2#3#4#5              {  \left(\begin{array}{ccccc}
                                     #1\\#2\\#3\\#4\\#5\end{array}
                              \right)  }
\def\eqabegin         {  \begin{eqnarray}  }
\def\eqaend           {  \end{eqnarray}  }
\def\nn               {  \nonumber  }
\def\bracetwo#1#2     {  \left\{ \begin{array}{l} #1 \\ #2 \end{array}
                         \right.  }
\def\bracetwocases#1#2#3#4  {   \left\{ \begin{array}{ll} #1 &
                                 \qquad #2 \\
                                 #3 & \qquad #4 \end{array} \right.  }
\def\bracebegin#1     {  \left\{ \begin{array}{#1}   }
\def\braceend         {  \end{array}\right.   }
\def\parn              {  \par\noindent }

\def\parmedskip        {  \par\medskip  }

\def\parbigskipn        {  \par\bigskip\noindent  }
\def\parmedskipn        {  \par\medskip\noindent  }
\def\parsmallskipn      {  \par\smallskip\noindent  }

\def\parag#1           {\paragraph{#1} \mbox{ }\parmedskip\noindent}
\def\boxit#1#2      {  \vbox{\hrule\hbox{ \hskip -4.1pt \vrule\kern3pt 

                     \vbox
                    {  \hsize #1 \strut\kern3pt #2 \kern3pt\strut  }
                       \kern3pt  \vrule} \hrule  } }
\def\centerbox#1#2  {  \mbox{  }\par\bigskip  \hfil \boxit{#1}{#2} \hfil
                       \par\bigskip\noindent }
\def\rightbox#1#2   {  \hfill\boxit{#1}{#2}  }
\def\leftbox#1#2    {  \boxit{#1}{#2}  }
\def\fullbox#1      {  \boxit{\textwidth}{#1}  }
\def\rightfigspacebegin  {  \par\noindent\begin{minipage}[t]{10cm}  }
\def\rightfigspaceend    {  \end{minipage}\par\noindent  }
\def\leftfigspacebegin   {  \par\noindent
                             \hspace*{10cm}\begin{minipage}[t]{6cm} }
\def\leftfigspaceend     {  \end{minipage}\par\noindent  }
\def\titleandfile#1#2   {  \begin{center}{\Large\bf #1}\end{center}
                            \par\begin{flushright} #2 \end{flushright}  }
\def\msection#1      {  \begin{center} \section{#1} \end{center}   }
\def\nsection#1      {  \let\boldface\bf \def\bf{} \section{#1}
                           \let\bf\boldface   }
\def\mnsection#1     {  \begin{center} \nsection{#1} \end{center}  }
\def\capsection#1    {  \let\boldface\bf \def\bf{\sc} \section{#1}
                           \let\bf\boldface   }
\def\mcapsection#1   {  \begin{center} \capsection{#1} \end{center} }

\def\sectionnumbering { \setcounter{equation}{0}
         \renewcommand{\theequation}{\arabic{section}.\arabic{equation}}}

\newcommand{\nullify}[1]{}
\def\ie{{\it i.e. }}

\def\fdot{{\dot{f}}}

\def\ehat{{\hat{e}}}

\def\ftil{{\tilde{f}}}

\def\abs#1{{|#1|}}
\def\bra#1{\langle #1 |}
\def\ket#1{| #1 \rangle}
\def\altil{{\tilde{\al}}}

\def\com#1#2{ \left[#1, #2\right]}

\def\util{{\tilde{u}}}

\def\adag{a^\dagger}
\def\atil{{\tilde{a}}}
\def\atildag{{\tilde{a}^\dagger}}

\def\btil{{\tilde{b}}}

\def\bebar{\bar{\beta}}

\def\xitil{\tilde{\xi}}

\def\papertitlepage{\baselineskip 3.5ex \thispagestyle{empty}}
\def\Title#1{\baselineskip 1cm \vspace{1.5cm}\begin{center}
 {\Large\bf #1} \end{center} 
\vspace{0.5cm}}
\def\Authors#1{\begin{center} {\it #1} \end{center}}
\def\Abstract{\vspace{1.0cm}\begin{center} {\large\bf Abstract} 
           \end{center} \par\bigskip}
\def\Komabanumber#1#2#3{\hfill \begin{minipage}{4.2cm} UT-Komaba #1
              \parn #2 
              \parn #3 \end{minipage}}
\renewcommand{\thefootnote}{\fnsymbol{footnote}}
\renewenvironment{thebibliography}{\pagebreak[3]\par\vspace{0.6em}
\begin{flushleft}{\large \bf References}\end{flushleft}
\vspace{-1.0em}

\begin{enumerate}\if@twocolumn\baselineskip=0.6em\itemsep -0.2em
\else\itemsep -0.2em\fi\labelsep 0.1em}{\end{enumerate}}
\begin{document}
\papertitlepage
\vspace*{0cm}
\Komabanumber{97-7}{hep-th/9705111}{May 1997}
\Title{Scattering of Quantized Dirichlet Particles } 
\vspace{1cm}
\Authors{{\sc 
  Yoichi Kazama
\footnote[3]{kazama@hep3.c.u-tokyo.ac.jp}
\\ }
\vskip 3ex
 Institute of Physics, University of Tokyo, \\
 Komaba, Meguro-ku, Tokyo 153 Japan \\
  }
\baselineskip .7cm
\Abstract
As a step toward satisfactory understanding of the quantum dynamics
 of Dirichlet \break (D-) particles, the amplitude for the basic 
  process describing the scattering of two quantized D-particles 
  is computed in bosonic string theory. The calucluation is performed  
 and cross-checked using  three different methods, 
 namely, (i) path integral, (ii) boundary state, and (iii) open-channel 
 operator  formalism.  The analysis is exact in $\al'$ and includes 
 the first order correction in the expansion with 
 respect to the acceleration of the D-particles. 
 The  resultant   Lorentz-invariant amplitude is capable of  describing
  general non-forward scattering with recoil effects fully taken into 
 account and it reproduces  the known result  for the special case of 
  forward scattering in the limit of infinitely large D-particle mass. 
 The expected form of the amplitude for the supersymmetric case is 
  also briefly discussed. 
\newpage
\baselineskip 0.65cm
\section{Introduction}  
 \sectionnumbering
\renewcommand{\thefootnote}{\arabic{footnote}}
A recent proposal on  microscopic formulation of  M theory
 \cite{BFSS,Towns9512} suggests that the 
Dirichlet (D-)particles, which  are thought to describe  solitonic 
 collective excitations of string\cite{Polch9510}, may actually  
play a much more fundamental role. In that  approach, at least 
in the 11 dimensional infinite momentum frame, D-particles are 
the basic  degrees of freedom and  extended objects such as 
 membranes, strings, etc., 
arise as collective excitations. A certain amount of evidence exists 
 in support of this aspiring conjecture, essentially 
in low energy domain.  
It is extremely important  to check if the details work out, especially 
 whether there would be higher derivative corrections to the proposed 
 action. To answer this question, obviously we need to know 
much more about the quantum interaction of D-particles. 
\par
The conjecture above  was  in part 
 motivated by the effective low energy description of the 
D-brane interactions  in the form of  
 super-Yang-Mills  theory on the worldvolume\cite{Witten9510}. 
Although conceptually 
 independent of  M-theory, some hints about the hidden 11 dimensional 
 feature have been uncovered\cite{KabPoul}\cite{DanFerSun}\cite{DKPS}, 
 suggesting that  extension of this approach is a viable road to 
   understanding of the underlying fundamental theory.  Another 
 intriguing  feature that appears in this approach 
  is  the apparent non-commutative 
 nature of the coordinates describing the D-branes. 
Although natural from the gauge theory point of view, its physical 
 meaning and how fundamental it actually is  yet to be clarified. 
Again more detailed understanding of the quantum dynamics 
 of D-branes is needed to make further progress.    
\par
With this background, we present in this article a computation of the 
 amplitude for scattering of two quantized D-particles with  finite mass
 in bosonic string theory.  The result, valid  when the accelerations 
 of the D-particles are  small in their mass scale, is  nevertheless 
 exact in $\al'$ 
 expansion and describes  general non-forward scattering  with 
 recoil effects fully included.  In the special case of  forward 
 scattering in the limit of infinitely heavy classical D-particles, 
the amplitude 
 correctly reduces to the known expression \cite{Bachas}. 
In fact, the structure of our result is formally quite similar 
 to the one for the special case so  that there exists
   a simple rule to go backwards, \ie to go from this 
 special case  to our general case of quantum D-particles. 
 Applying  this rule, we will write down the result 
 expected in the superstring 
case\footnote{In superstring case, satisfactory  analysis should 
 include the processes involving fermionic superpartners of 
(bosonic) D-particles  and this is yet to be performed.}as well
 in the final section.  Our result should 
 serve as a firm knowledge to be checked against in any proposal 
 for the fundamental microscopic theory containing D-particles. 
\par
The actual calculations are perfomed and cross-checked 
 by three different methods, namely (i) the path integral, 
(ii) the boundary state,  and (iii) the operator method in the open 
 string channel.  The emphasis is on the path integral method, which
 is conceptually most complete in formulating the problem and 
 hence is capable of computing the corrections in powers 
 of acceleration  systematically.  It  is a non-trivial 
  extension of the formalism that we developed recently\cite{paperI} for 
 the scattering of closed string states from a quantized D-particle. 
We will see that all the details work out consistently , including 
 the first order correction\footnote{In this article, 
 \lq\lq order" invariably refers to the one in the power expansion 
 with respect to the acceleration of the D-particles.}
 that  we will take into account  in this article. 
As it will become clear later, the other two methods, although not 
 suitable for computing corrections in powers of acceleration, have 
 their own advantages. The boundary state method turned out to be 
 most efficient for the purpose of computing the lowest order 
amplitude. On the other hand, the  operator method 
 reveals a deformation of the spectrum of an unusual type 
 in the open string  channel and the transitions among these 
 excitations. Deeper understanding of this phenomena may be quite 
 important in connection with the effective gauge theory.   
\par
 In order to explain three different 
 methods,  each of which  contains subtleties both technical and
 conceptual, in a reasonably  self-contained manner, this article 
 has become somewhat long. However, we have organized it in 
 such a way that a hurried reader, if so desires,  may consult section 2 on 
 the path integral method, which by itself is complete, and skip 
 sections 3 and 4 describing the other two methods. 
 Here is the content of the rest of the article to follow:
\par
Section 2. is devoted to the path integral method.  In subsection 2.1, 
 after describing how D0-D0 system should be  characterized,
  we carefully analyze the boundary conditions when the 
 D-particles follow arbitrary trajectories. They are then expressed 
 in a  suitable orthonormal coordinate system, to be 
 used throughout the paper.  Path integral over the string 
 coordinate itself is performed in subsection 2.2. The  
 complicated boundary conditions are implemented  by 
 devising a useful trick, and the lowest order amplitude as well as the 
 Green's functions needed for the calculation of the corrections are
 computed. In subsection 2.3  the corrections, 
 to first order in the acceleration, 
 which arise from the extendedness of the boundaries are computed.   
The divergence produced in this process is consistently  absorbed 
 by the renormalization of the trajectories. Finally, in subsection 2.4
 we quantize the D-particle trajectories themselves and obtain the 
 desired quantum amplitude. It is then compared with the known result 
 in the special limit already mentioned above. \par
In section 3, we briefly describe how the lowest order amplitude is 
 reproduced using the boundary state representation of the interaction
 vertex developed previously\cite{Ishibashi}\cite{paperI}. 
 After writing down the appropriate vertex states  in 3.1, 
 we sketch the actual computation in subsection 3.2. 
 A powerful normal-ordering formula is developed (a proof is sketched 
 in the Appendix) and used to calculate the Green's functions as well 
 as the amplitude. \par
The computations performed in section 2 and 3 are essentially from 
 the closed string channel. In section 4, we develop the 
operator formalism in the complimentary open string channel. 
First in 4.1, we make the modular transformation of the amplitude 
previously computed and obtain the form to be reproduced in the open
 channel. Upon performing the operator quantization in 4.2, we will 
find that the structure of the Hilbert space including the energy 
spectrum is rather unusual. We will give a careful analysis of this 
 structure, give physical interpretation, and compute the relevant 
 trace in a reliable way. The end result is exactly the one expected. 
\par
Finally, in section 5, we make several  important remarks that 
 emerge from our work. One among them is about   
 the expected form of the amplitude in superstring theory. 
 We will spell out how, by applying the aforementioned rule, 
  the result of \cite{Bachas} for the special case 
 can be promoted to the amplitude for  quantized D-particles. 
 Another remark concerns the existence of infinite number 
 of diagrams contributing to the scattering 
 in the high energy regime even for small string coupling. 
\section{Path Integral Approach}
\setcounter{equation}{0}
\subsection{Characterization of D0-D0 system}
\noindent{\bf The setup} \parsmallskipn
A D-particle (D0-brane) is a point-like object which can emit and absorb 
a closed string. It does so in such a way that, when looked at 
 from the (dual) open string channel, the ends of the open string 
 lie somewhere on its worldline. Thus the simplest diagram describing 
 two such D-particles interacting with each other looks like 
 the one dipicted in Fig.1. 
\begin{center} 
\epsfxsize=5cm
\quad\epsfbox{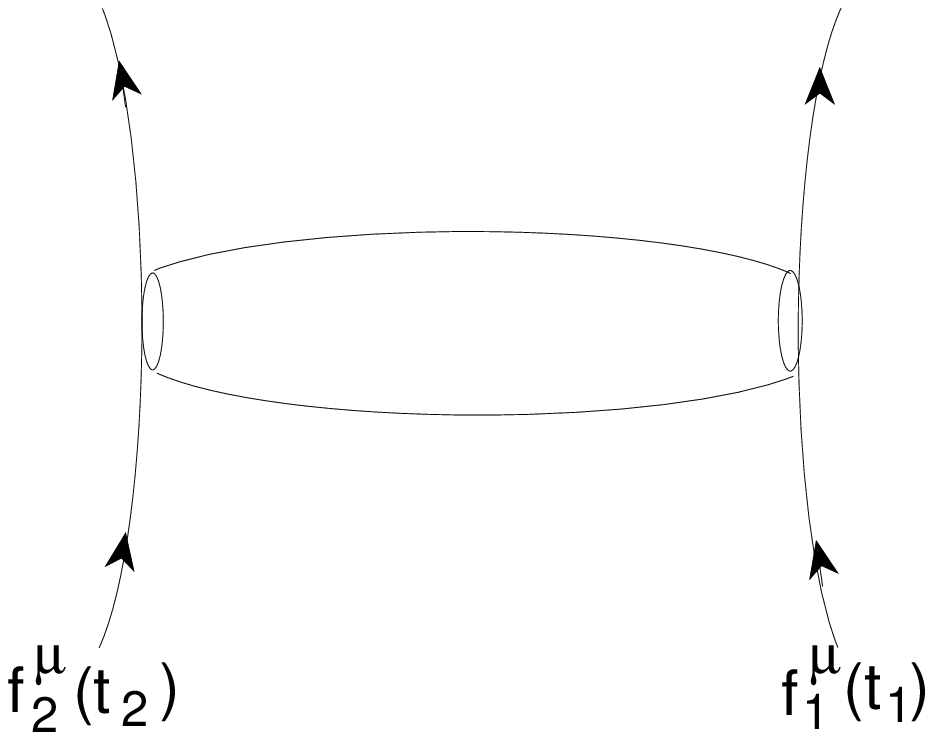} \\
{\small {\bf Fig.1} \quad Basic process for D0-D0 scattering.}
\end{center}
\parn
%
%
In the open string language, the relevant 
topology of the string  worldsheet is an annulus $\Sigma$, 
 which we take to be one with  the inner radius $R_1$ and the  outer 
radius $R_2$. The boundary circle at each end will be  parametrized 
 by the angle $\theta$. We take the inner boundary to be mapped 
 onto the worldline of  a D-particle,  parametrized by 
$f^\mu_1(t_1)$, while the outer boundary is embedded into the 
 worldline of the other D-particle, described by $f^\mu_2(t_2)$. 
Until we come to the subsection 2.4, where we quantize the D-particles, 
we take $f^\mu_i(t_i),\, i=1,2$ to be arbitrary yet fixed time-like
 trajectories\footnote{There will be a slight condition required 
 on $f^\mu_i(t_i)$ for consistency later.} . 
 As the ends of the open string may 
 terminate anywhere on the worldlines, the Lorentz covariant 
 constraints at the boundaries should be of the form \cite{Leigh}
\eqabegin
X^\mu(r=R_i,\theta) &=& f_i^\mu(t_i(\theta)) \comma 
\label{bcst}
\eqaend
where $X^\mu(r,\theta)$ denote the open string variables in the 
 worldsheet polar coordinates and $t_i(\theta)$ are 
 arbitrary functions describing the embedding. 
 This means that in the path integral formulation we develop  
in this section, we must integrate over $X^\mu(r,\theta)$
 in the bulk and over $t_i(\theta)$ at the boundaries. Thus the 
 relevant amplitude is 
\eqabegin
 \calV(f_1,f_2) &=& \int \calD X^\mu(r,\theta) 
  \prod_i \calD t_i(\theta) \prod_i \delta (X^\mu(R_i,\theta)
 -f_i^\mu(t_i(\theta)) e^{-S\left[X\right]} \comma 
\eqaend
where 
\eqabegin
 S\left[X\right] &=& {1\over 4\pi \al'} \int _\Sigma 
 d^2z \del_\al X^\mu \del_\al X_\mu 
\eqaend
is the open string action\footnote{We use Euclidean worldsheet
 and Minkowski target space with the metric 
 $\eta_{\mu\nu} = {\rm diag}\ (-,+,+,\cdots, +)$.}. \par
\shead{Scheme for $t_i(\theta)$-integration}
To give a feeling for how we will compute the amplitude above, 
 we  need to briefly recall the scheme for 
$t_i(\theta)$-integration explained in \cite{paperI} (for the 
 single D-particle case). 
 We first split the integral over $t_i(\theta)$ into 
 the one over the $\theta$-independent mode, denoted by $t_i$, and the 
 rest over the non-constant modes,  and then, to retain 
 general covariance,  expand 
 $f^\mu_i(t_i(\theta))$ in terms of the geodesic normal coordinate 
$\zeta_i(\theta)$ around $f_i^\mu(t_i)$.  It can be written in the 
 form 
\eqabegin
f^\mu_i(t_i(\theta)) &=& f_i^\mu(t_i) + \fdot_i^\mu(t_i) \zeta_i(\theta) 
 + \Omega_i^\mu(\theta) \comma \label{geoexp} \\
\where \Omega_i^\mu(\theta) &\equiv & 
 \half K^\mu_i\zeta_i(\theta)^2 \nn\\
&& + {1\over 3!} \Biggl(
 -{\fdot_i^\mu \over h_i}K_i^2 -{3\over 2}{\dot{h}_i \over h_i}K_i^\mu
 +P^{\mu\nu}_i \del_t^3 f_{i\nu} \Biggr) \zeta_i(\theta)^3 + \cdots
\period
\eqaend
Here, a dot stands for a  $t$-derivative 
 and $h_i(t_i)$, $K_i^\mu(t_i)$
and  $P_i^{\mu\nu}(t_i)$ are, respectively, the one-dimensional 
 induced metric on the trajectory, the extrinsic curvature and 
 a projection operator normal to the trajectory. They are 
 given by 
\eqabegin
h_i &\equiv & \fdot_i^\mu \fdot_{i\mu}\comma \\
K_i^\mu &\equiv & \ddot{f}_i^\mu -\half{\dot{h}_i\over h_i} 
 \fdot_i^\mu = 
P^{\mu\nu}_i\ddot{f}_{i\nu}\comma  \label{kmudef} \\
P^{\mu\nu}_i&\equiv & \eta^{\mu\nu} -h_i^{\mu\nu}\comma\label{projp} \\ 
 h^{\mu\nu}_i &\equiv&{\fdot_i^\mu\fdot_i^\nu \over h_i}\comma
\label{projh}
\eqaend
where we have also introduced the projection operator $h_i^{\mu\nu}$ 
along the trajectory. The integration over $\zeta_i(\theta)$ 
 cannot, unfortunately, be performed in an exact manner. 
As will be explained in detail in subsection 2.3, we will 
 organize it perturbatively with respect to the order of 
 $t$-derivatives. This expansion scheme 
 is expected to be reliable when the accelerations of the D-particles 
 are small compared with the D-particle mass scale. Diagramatically, 
 it corresponds to the picture (see Fig.2) that the dominant 
 interaction occurs at a point on the worldline and the corrections 
 due to the extendedness of the boundaries are subsequently taken 
 into account by $\zeta_i(\theta)$-integrations. 
\begin{center} 
\epsfxsize=12cm
\quad\epsfbox{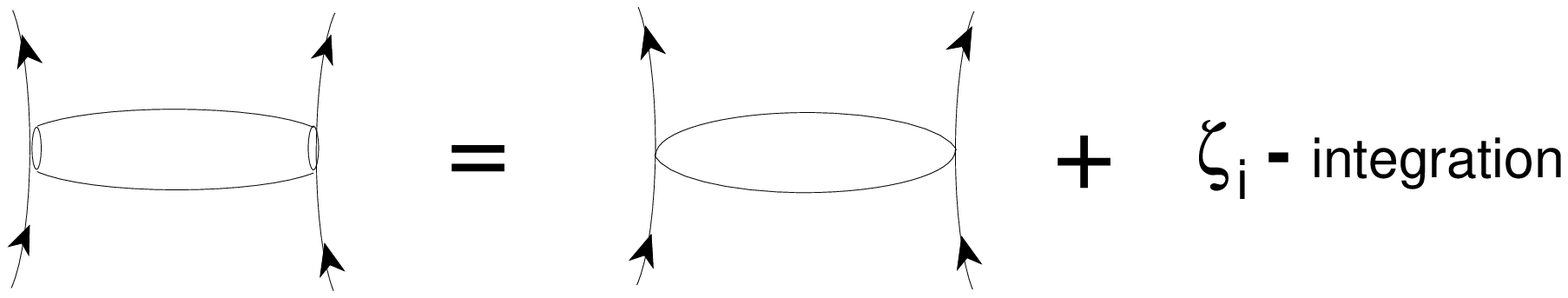} \\
{\small {\bf Fig.2} \quad Diagram showing our approximation scheme.}
\end{center}
\parn
%
%

\shead{Splitting of string coordinate}
To perform the integration over $X^\mu(r,\theta)$, it is convenient
 to first split it into  the $\theta$-independent part   $X^\mu_0(r)$ 
 and the rest called $\xitil^\mu(r,\theta)$:
\eqabegin
X^\mu(r,\theta)&=& X^\mu_0(r)+\xitil^\mu(r,\theta)\comma \\
\int d\theta \xitil^\mu(r,\theta) &=& 0\period
\eqaend
Then the action separates without surface term into 
\eqabegin
 S &=& {1\over 2\al'}\int {dr\over r} (r\del_r X_0(r))^2 
 + {1\over 4\pi \al'}\int d^2z (\del_\al \xitil)^2 \period \label{action2}
\eqaend
The constraints at the boundaries now read
\eqabegin
X_0^\mu(R_i) &=& f^\mu(t_i) +  \Omega^\mu_{i,0} \comma \label{cstxz}\\
\xitil^\mu(R_i,\theta) &=& \fdot_i^\mu(t_i) \zeta_i(\theta) 
 + \tilde{\Omega}_i^\mu(\theta) \comma  \label{cstxitil}\\
\where \Omega^\mu_{i,0}&\equiv & \int{d\theta \over 2\pi} 
\Omega_i^\mu(\theta)\comma \label{omzero} \\
\tilde{\Omega}^\mu_i (\theta) &\equiv & \Omega^\mu_i(\theta) 
 -\Omega^\mu_{i,0} \period \label{omtil}
\eqaend
Note that by definition $\zeta_i(\theta)$ itself has no constant part, 
 namely, $\int d\theta \zeta_i(\theta)=0$.  
\shead{Boundary conditions} 
We now formulate the boundary conditions 
that follow from the consistency under  the  variation of the action
 (\ref{action2}). As we have to take into account the  
 constraints (\ref{cstxz}) and (\ref{cstxitil}) 
 already existing at the boundary, the analysis is 
somewhat involved.
It is easy to see that  the boundary conditions compatible with
 the equations of  motion  in the bulk are
\eqabegin
\del_r X_0(R_i)\cdot  \delta X_0(R_i) &=& 0 \comma \label{delx0}\\
\del_r \xitil(R_i, \theta) \cdot \delta  \xitil(R_i, \theta) &=& 0\period
\label{delxitil}
\eqaend
As has already been stated, we will treat $\calO(\zeta^2)$ terms 
 in (\ref{geoexp}) (which involve derivatives higher than the second) 
 perturbatively. Therefore,  these conditions need be imposed 
 up to $O(\zeta)$. 
As for $X^\mu_0$, the constraint (\ref{cstxz}) at the boundary says that 
the only possible variation up to this order  is of the form 
 $\delta X^\mu_0(R_i)=\fdot_i^\mu(t_i)\delta t_i$, \ie in the 
 direction of the trajectory.  Similarly, (\ref{cstxitil}) dictates 
 that (since $\zeta_i(\theta)$ is considered small)
   possible variation of $\xitil^\mu(R_i,\theta)$ is also  along the  
 trajectory. Thus, using the 
 projection operators introduced in (\ref{projp}) and  (\ref{projh}), we 
 can write the conditions as  
\eqabegin
\delta X_0^\mu(R_i) P_{\mu\nu,i} &=& 0 \comma \qquad 
\del_r X_0^\mu(R_i) h_{\mu\nu,i} = 0 \comma \label{bcxz}\\
\delta \xitil^\mu(R_i,\theta ) P_{\mu\nu,i} &=& 0 \comma\qquad 
\del_r \xitil^\mu(R_i,\theta) h_{\mu\nu,i} = 0 \period \label{bcxitil}
\eqaend
To see the content of the conditions for $X^\mu_0(R_i)$ in more 
 detail, it is useful to 
 further split  $X^\mu_0(r)$ into the classical part $X^\mu_{cl}(r)$
  that satisfies  $\del^2X^\mu_{cl}(r)=0$,  $ X^\mu_{cl}(R_i) 
=f_i^\mu(t_i)$, and the quantum part. The desired splitting 
 can be written as  
\eqabegin
X_0^\mu(r)&=& X^\mu_{cl}(r) +{x^\mu_0(r)\over \sqrt{4\pi}} \comma \\
X^\mu_{cl}(r) &=& {1\over s }
 \left( f^\mu_1(t_1)\ln{b \over r} + f^\mu_2(t_2)\ln{r \over a} \right)
\comma \label{xclassical} \\
\where s &\equiv & \ln{b\over a} \period \label{defs}
\eqaend
Putting this into (\ref{bcxz}), we get 
\eqabegin
0 &=& \delta x^\mu_0(R_i) P_{\mu\nu,i} \comma \\
0&=& {1\over sR_i}(f_2(t_2) -f_1(t_1))^\mu h_{\mu\nu,i}
 + {1\over \sqrt{4\pi}}\del_r x_0^\mu(R_i) h_{\mu\nu,i} \period
\eqaend
The first of these equations simply says that $x^\mu_0$ should 
 satisfy the Dirichlet condition for the direction transverse to the 
 trajectory at each end. The second, on the other hand, dictates 
 that we should impose the Neumann condition for $x_0^\mu$ in 
 the tangential direction and at the same time demand, for consistency, 
\eqabegin
(f_2(t_2)-f_1(t_1))_\mu \fdot_1^\mu(t_1) &=& 
(f_2(t_2)-f_1(t_1))_\mu \fdot_2^\mu(t_2)=0\period
\eqaend
 These additional 
 conditions may look odd at first sight, but their physical meaning 
is clear: The quantity $f^\mu_2(t_2)-f^\mu_1(t_1)$ is conjugate 
 to the momentum transfer for the D-particles 
 and the condition simply says that 
 to $\calO(\zeta)$ there should not be any  momentum transfer along the 
 trajectory so that the D-particles should be able to move  freely along 
 the trajectory\footnote{Remember that the effect of the 
 extrinsic curvature starts at $\calO(\zeta^2)$.}. 
 We shall see later  in sec.~3 that  such conditions arise 
  also from the requirement of BRST invariance for the boundary state 
 representation of the interaction vertex.  
 \par
The remaining analysis for $\xitil^\mu$ is straightforward and we get
 the same type of boundary conditions as for $x^\mu_0$. So if we 
 define the total quantum fluctuation $\xi^\mu(r,\theta)$ by 
\eqabegin
 \xi^\mu(r,\theta) &=& {x^\mu_0(r) \over \sqrt{4\pi}} 
+\xitil^\mu(r,\theta) \comma 
\eqaend
we can summarize the boundary conditions as 
\eqabegin
P^{\mu\nu}_i \xi_\nu(R_i,\theta) &=& 0 \comma \label{dirbc}\\
h^{\mu\nu}_i \del_r \xi_\nu(R_i,\theta) &=& 0 \period\label{neubc}
\eqaend
%
\shead{Orthonormal basis}
The path integral over the quantum fluctuations $\xi^\mu(r,\theta)$ 
 will be facilitated if we set up a suitable orthonormal 
basis. In the one particle case\cite{paperI}, 
there is  a unique natural such frame, which is 
 spanned by the unit vector $u^\mu=\fdot^\mu/\sqrt{-\fdot^2}$ 
 tangent to the trajectory and the remaining $D-1$ transverse,
 mutually orthogonal unit vectors. Then the boundary 
 conditions are neatly separated into the Neumann in the tangent and 
 the Dirichlet in the transverse directions.  \par
For the case at hand, however, we have {\it two ends} and the Neumann 
directions at these end points are {\it different}. Thus, 
 we may consider two possible orthonormal frames, to be defined below. 
\par
Since $\fdot_i^\mu$ are expected to be time-like\footnote{When 
 the D-particles are put on-shell after quantization, this will be 
automatic.}, let us first introduce the time-like unit vectors $u_i$ 
 defined by
\eqabegin 
 u_i &\equiv & {\fdot_i^\mu \over \sqrt{-h_i}}\comma \qquad 
u_i^2 = -1 \period
\eqaend
In the generic case, $u_1$ and $u_2$ are non-degenerate and 
define a plane, which we call \lq\lq the trajectory plane". 
It is easy to show that the  $D$-dimensional Lorentizan inner product 
$u_1\cdot u_2$ takes the values in the range $\infty \ge 
 -u_1\cdot u_2 \ge  1$, where the equality on the right  holds if and only if 
the their spatial parts agree, \ie if $\vec{u}_1(t_1)=\vec{u}_2(t_2)$.
 Therefore we can parametrize $-u_1\cdot u_2$ in the form 
\eqabegin
-u_1\cdot u_2 &=& \cosh \chi\comma 
\eqaend
where $\chi$ is real and positive (by convention)
 and it depends on $\fdot_i$ symmetrically. {\it This parameter $\chi$ 
 will be of utmost importance as it will be shown, in subsection 2.4, to 
carry the essential information of the scattering.}  Its physical meaning 
 will also be spelled out there. 
\par
Now we can define the unit-normalized space-like vectors
 $\util_2$ and $\util_1$ in the trajectory plane, which are
 orthogonal to $u_1$ and $u_2$ respectively:
\eqabegin
 \util_2 &=& {1\over \sinh\chi} (u_2 -u_1\cosh\chi) \comma 
\qquad \util_2^2 = 1 \comma \qquad u_1\cdot \util_2=0\comma \\
 \util_1 &=& {1\over \sinh\chi} (u_1 -u_2\cosh\chi) \comma 
\qquad \util_1^2 = 1 \comma \qquad u_2\cdot \util_1=0\period 
\eqaend
Either $\left\{u_1,\util_2\right\}$ or
 $\left\{u_2,\util_1\right\}$ will do, but we 
 will choose the former for our basis in the trajectory plane. 
The latter then is expressed in terms of the former in the form  
\eqabegin
u_2 &=& u_1\cosh\chi + \util_2\sinh\chi\comma \label{u2}\\
\util_1 &=& -(u_1\sinh\chi +\util_2\cosh\chi) \period \label{util1}
\eqaend
By appending $D-2$ space-like orthonormal vectors 
$u_I$ which span  the space transverse to the trajectory plane, 
we complete our basis as 
\eqabegin
\left\{\ehat^\mu_A\right\} 
&=& \left\{ u_1, \util_2, u_I\right\} \comma \quad 
A =(i,I) \comma\quad  \ i=1,2\comma \ 
I = 3,4,\ldots, D\period \label{stdbasis}
\eqaend
Accordingly, we will define the components of $\xi^\mu$ in 
 this frame as 
\eqabegin
 \xi_A &\equiv & \ehat^\mu_A \xi_\mu \comma 
\eqaend
that is,  $\xi_1 \equiv  u_1\cdot \xi$, 
 $\xi_2 \equiv \util_2 \cdot \xi$  and  
 $\xi_I \equiv u_I\cdot \xi$.
With this notation, we have 
\eqabegin
a_\mu b^\mu &=& a_Ab^A = a_A \eta^{AB}b_{BA}= a^A\eta_{AB}b^B\comma \\
\eta^{AB} &=& \eta_{AB}= (-,+,+,\cdots, +) \comma 
\eqaend
for arbitrary Lorentz vectors $a_\mu,b_\mu$. 
\par
It is now a simple matter to express the boundary conditions 
  (\ref{dirbc}) and (\ref{neubc})  in our basis: 
\eqabegin
&& \del_r\xi_1(R_1,\theta) = 0 \comma  \label{bca1} \\
&& \xi_2(R_1,\theta)=0 \comma \label{bca2} \\
&& \xi_1(R_2,\theta) \sinh\chi +\xi_2(R_2,\theta) \cosh\chi=0
\comma  \label{bcb1} \\
&& \del_r \xi_1(R_2,\theta)\cosh\chi + \del_r\xi_2(R_2,\theta)\sinh\chi 
\comma \label{bcb2} \\
&&  \xi_I(R_1,\theta) = \xi_I(R_2,\theta) =0   \label{bci} \period
\eqaend
Note that these conditions are similar to but more involved than 
 the  ones that occur in the case of  usual open string in a 
constant electromagnetic field \cite{CallanII,BachasPorrati,Nesterenko}
 or in the case of  D-particles moving in the same 
 directions\cite{Bachas}.
  In the present case, the two ends couple to background 
trajectories which point in different directions 
  and  hence  the boundary conditions for $\xi_1$ 
and $\xi_2$ cannot be fully disentangled. 
\subsection{Path integral over the string coordinates}
\shead{Preliminary}
We shall now perform the integration over $\xi^\mu(z)$. 
Taking into account the boundary constraints 
(\ref{cstxz}), (\ref{cstxitil}), the action can be written 
 as 
\eqabegin
S &=& S_{cl} +S_{cl-q} + S_{q} + S_{cst} \comma \\
S_{cl} &=& {1\over 2\al' s}(f_2-f_1)_\perp^2 \comma \label{scl} \\
S_{cl-q} &=& {1\over \al's}(f_2-f_1)_\perp^\mu (\Omega_{2,0}
 -\Omega_{1,0})_\mu \comma \label{sclq} \\
S_{q} &=& {1\over 4\pi \al'}\int d^2z (\del_\al\xi)^2 \comma \\
S_{cst} &=& -i\int d\theta \sum_i \nu_\mu(R_i,\theta)
 \left( \xitil^\mu(R_i,\theta) -\fdot^\mu_i(t_i)\zeta_i(\theta)
 -\tilde{\Omega}^\mu_i(\theta)\right) \comma \label{scst}
\eqaend
where $(f_2-f_1)_\perp$ denotes the components orthogonal to 
 the trajectory plane.  In (\ref{scst}) 
 we have introduced sources $\nu_\mu(R_i,\theta)$,  
{\it without  constant parts},   at 
 the boundaries in $S_{cst}$, which will be integrated to produce 
 $\delta$-functions enforcing the constraints (\ref{cstxitil}). 
 ((\ref{cstxz}) is already used in (\ref{sclq}).) 
 
The path integral over   
 $\xi^\mu(z)$  then can be written in the form 
\eqabegin
\calV_{\xi} &=&  \int \calD \xi^\mu e^{E_\xi} \comma \\
 E_\xi&=& {1\over 4\pi \al'} \int d^2z \xi(z)\cdot \del^2 \xi(z) + 
i\int d^2 z \xi(z) \cdot J(z) \period
\eqaend
Although the source $J(z)$, representing $\nu_\mu(R_i,\theta)$, 
 is active only on the boundaries, we shall, until appropriate time,
 take it to be a general function of $z$. 
This will allow us to compute the requisite 
 Green's functions on the annulus satisfying the rather complicated 
 boundary conditions imposed on $\xi (z)$. 
 Also for a while we shall suppress the component indices for 
 $\xi$ and $J$ unless needed. \par
In the following, it will be  convenient to switch from 
the variable $r$ to a more natural proper-time variable $\rho$ 
 defined by
\eqabegin
 \rho &\equiv&  \ln{r \over R_1} \comma \qquad 
 0 \le \rho \le s \equiv \ln{R_2\over R_1} \period 
\eqaend
The Laplacian and the integration measure  take the 
 form, $ \del^2 = R_1^{-2}e^{-2\rho}(\del_\rho^2+\del_\theta^2)$ 
 and $\int d^2z = R_1^2 \int_0^{2\pi}d\theta \int _0^s d\rho e^{2\rho}$.
\par
We now expand various quantities in double Fourier series 
in $\theta$ and $\rho$ and then integrate over the modes. 
First expand $\xi$ and $J$ in angular Fourier series:
\eqabegin
 \xi(\rho,\theta) &=& \sum_n {1\over \sqrt{2}} (x_n(\rho) + ix'_n(\rho))
 {e^{in\theta} \over \sqrt{2\pi}} \comma \\
J(\rho,\theta) &=&   \sum_n {1\over \sqrt{2}} (j_n(\rho) + ij'_n(\rho))
 {e^{in\theta} \over \sqrt{2\pi}} \period
\eqaend
From the reality conditions, 
$x_n$ and $x'_n$ must be real and satisfy $x_{-n} = x_n$, 
$x'_{-n} =-x'_n$. So the 
independent integration variables are $x_n$ and $x'_n$ for $n\ge 1$
 and $x_0$. Then  the exponent $E_\xi$ becomes 
\eqabegin
E_\xi &=& -{1\over 4\pi\al'} \int _0^s d\rho
\left[ \sum_{n=0}^\infty 
  x_n(-\del_\rho^2 +n^2)x_n 
+ \sum_{n\ge 1} x'_n(-\del_\rho^2 +n^2)x'_n\right]\nn\\
&& + iR_1^2 \int _0^s d\rho e^{2\rho} \left[ \sum_{n=0}^\infty
 x_n j_n + \sum_{n\ge 1} x'_n j'_n \right]  \label{exi} \period
\eqaend
Since the primed system is identical to the unprimed 
(except for $n=0$ part), we will hereafter exhibit the unprimed part
 only. 
\par
Next we wish to make a Fourier expansion of $x_n(\rho)$  with 
 respect to $\rho$ in the interval  $\left[0,s\right]$. 
 Here we must take into account the boundary 
 conditions (\ref{bca1})$\sim$(\ref{bci}), which in terms of 
 $x_n(\rho)$'s read 
\eqabegin
&& \del_\rho x_{1,n}(0)= 0 \comma \label{bcxa1} \\
&& x_{2,n}(0)= 0 \comma \label{bcxa2} \\ 
&& x_{1,n}(s) \sinh\chi + x_{2,n}(s)\cosh\chi  =0 
\comma \label{bcxb1} \\
&& \del_\rho x_{1,n}(s) \cosh\chi + \del_\rho x_{2,n}(s)\sinh\chi 
 =0 \comma \label{bcxb2} \\
&& x_{I,n}(0) =x_{I,n}(s) =0 \period \label{bcxi} 
\eqaend
The expansion for $x_{I,n}$, which satisfies the usual Dirichlet 
condition at both ends, is standard. But finding the appropriate 
Fourier expansions for $x_{1,n}(\rho)$ and $x_{2,n}(\rho)$,  obeying 
complicated boundary conditions,  is rather difficult.
 To circumvent this, we will use the following 
 trick. The idea is to first choose some appropriate complete basis, 
which satisfies a certain  boundary condition, 
   and then realize the actual boundary conditions 
 (\ref{bcxa1})  $\sim$ (\ref{bcxi}) by introducing extra source terms 
 $\Delta j_n(\rho)$, to be discussed in detail later.  For the moment, 
 we include them in $j_n(\rho)$ and proceed. 
\par
To find  a convenient as well as consistent Fourier  basis, 
 we must take into account  that the conditions 
 (\ref{bcxa1})  $\sim$ (\ref{bcxi}) are not periodic in the 
 interval  $\left[0,s\right]$. This suggests  that  we should 
 extend the functions into  the enlarged interval $\left[-s,s\right]$
 as {\it even}  functions and then 
 make the usual Fourier expansion. It reads   
\eqabegin
 x_n(\rho) &=& {1\over \sqrt{s}} \left( a_{n,0} 
 + \sqrt{2} \sum_{m=1}^\infty  a_{n,m} \cos k_m\rho \right)
 \label{neuexp} \comma \\
\where k_m &=& {\pi m \over s} \period
\eqaend
Note that this automatically satisfies the {\it Neumann} condition at 
 the ends of the {\it original} interval, \ie at $\rho=0,s$, 
 which is still consistent with the true boundary conditions. 
We will use this as our base and append corrections to realize the 
 true boundary conditions. \par
So we first compute the path integral for this basic case. 
 Substituting (\ref{neuexp})  
 into (\ref{exi}),   we get  for the mode $n$ 
(with the omission of the primed modes),
\eqabegin
E_{\xi,n} &=&  -{n^2\over 4\pi \al'} a_{n,0}^2 -
{1\over 4\pi\al'} \sum_{m=1}^\infty
 (k_m^2 +n^2) a_{n,m}^2  \nn\\
&& \qquad +ia^2\left( a_{n,0}j_{n,0} 
 + \sum_{m=1}^\infty a_{n,m} j_{n,m} \right)\comma 
\eqaend
where
\eqabegin
&&  j_{n,0} = {1\over \sqrt{s}} \int _0^s d\rho e^{2\rho} 
 j_n(\rho) \comma \label{jn0}\\
&& j_{n,m} = \sqrt{{2\over s}}  \int _0^s d\rho e^{2\rho}j_n(\rho)
\cos k_m\rho \period \label{jnm}
\eqaend
The Gaussian integrations over the modes $a_{n,0}$ and 
 $a_{n,m}$ are now trivial and it produces\footnote{Here
 and hereafter, we will not be concerned with the overall 
constant. If desired, it can be most easily obtained by the open 
 channel operator method, to be discussed in section 4.} 
\eqabegin
 \int  da_{n,0} \prod_{m=1}^\infty da_m 
 \, e^{E_{\xi,n}} &\sim & D^{(N)}_n(s)\,   e^{E(j)_n}\comma 
\eqaend
where 
\eqabegin
 D^{(N)}_n(s) &\sim & {1\over n} 
 \prod_{m=1}^\infty \left( k_m^2+n^2\right)^{-1/2}
 = {1\over n} \left({2\sinh ns \over n}\right)^{-1/2}\comma \\
E(j )_n&=&  -\pi \al' R_1^4\left[
 {1\over n^2} (j_{n,0})^2+\sum_{m=1}^\infty
{1 \over k_m^2+n^2} (j_{n,m})^2\right] \nn\\
&=& -\pi \al' R_1^4 \int_0^sd\rho \int_0^s d\rho' 
e^{2(\rho+\rho')} j_n(\rho) N_n(\rho,\rho')
 j_n(\rho') \period\label{expj}
\eqaend
Here, $N_n(\rho,\rho')$ is the Neumann function for the operator 
$-\del^2_\rho +n^2$ given by 
\eqabegin
N_n(\rho,\rho') &=& {2\over s} \left( {1\over 2n^2} +\sum_{m=1}^\infty
 {\cos k_m\rho \cos k_m\rho'\over k_m^2 +n^2}\right) \nn\\
&=& {1\over 2n \sinh ns } \left( \cosh n(s-(\rho+\rho')
 +\cosh n(s-|\rho -\rho'|) \right) 
\eqaend
with normalization $
(-\del^2_\rho +n^2)N_n(\rho, \rho') = \delta(\rho-\rho')$.
These expressions are derived with the aid of  the  formulae 
\eqabegin
\prod_{m=1}^\infty \left( 1+{y^2 \over m^2}\right) &=& {\sinh \pi y \over 
 \pi y}\comma  \\
 \sum_{m=1}^\infty {\cos mx \over m^2 +a^2} &=& 
 {\pi \over 2a}{\cosh a(\pi -x) \over \sinh a\pi} -{1\over 2a^2}\comma 
\quad (0\le x \le 2\pi) \period 
\eqaend
\par
We must now take the product over $n$. Due to the presence of the 
 constant mode $a_{0,0}$, the contribution for $n=0$ is divergent. 
As we shall see shortly, for the actual  boundary conditions  for our 
 problem, such divergences will cancel. Thus, for now  
 we  write $n=0$ as  $n=n_0$,  where it remains explicitly,  
  and  later check the cancellation.  
The front factor then becomes (including the contribution of 
 $x'_n(\rho)$), 
\eqabegin
D^{(N)}(s) &=& D_{n_0}(s) \prod_{n=1}^\infty  D_n^{(N)}(s)^2 \nn\\
&\sim & {s^{-1/2}\over n_0 } \prod_{n=1}^\infty 
\left({\sinh ns\over n}\right)^{-1} 
\sim  {s^{-1/2}\over n_0 } \eta(is/\pi)^{-1} \comma \label{dneu}
\eqaend
where  we  used a formula  
 $ \prod_{n=1}^\infty \sinh ns = \sqrt{2} \eta(is/\pi) $
  yielding the Dedekind $\eta$-function
$\eta(x) \equiv e^{i\pi x/12}\prod_{n=1}^\infty (1-e^{2\pi inx})$, 
and  employed 
 the standard $\zeta$-function regularization, such as 
 $ \prod_{n=1}^\infty A = A^{-1/2}$ and  $\prod_{n=1}^\infty n = \sqrt{2\pi} $, 
 where appropriate. 
\shead{Transverse sector}
Let us now describe how one can realize the true boundary
 conditions (\ref{bcxa1})  $\sim$ (\ref{bcxi}) by introducing 
 extra source terms. \par
We begin with the case of transverse components $x_{I,n}(\rho)$, 
which satisfy the  Dirichlet  conditions at both ends, 
 namely $x_{I,n}(0)=x_{I,n}(s)=0$.  In this case,  we need not, of course, 
 use any trick. Omitting the subscript $I$, we may simply expand $x_n(\rho)$ in the Fourier sine series 
\eqabegin
 x_n(\rho) &=& \sqrt{{2\over s}}\, \sum_{m=1}^\infty 
 d_{n,m} \sin k_m\rho \comma  
\eqaend
and go through the procedure similar to the previous case. One easily
 finds   
\eqabegin
D^{(D)}(s) &\sim & s^{-1/2} \eta(is/\pi)^{-1} \label{ddir}
\eqaend
for the front determinant factor,  and  in the exponent  of (\ref{expj}) 
 the Neumann function is replaced  by the Dirichlet function 
\eqabegin
D_n(\rho,\rho') &=& {1\over 2n \sinh ns } \left(- \cosh n(s-(\rho+\rho')
 +\cosh n(s-|\rho -\rho'|) \right)\period
\eqaend
This suffices for the Dirichlet case. 
However, in preparation for handling the more difficult case of 
 $\xi_1$-$\xi_2$  system,
  it is  important to check if  the same result can be reproduced 
 by our more versatile trick, to be described below.  \par 
In terms of the modes defined in (\ref{neuexp}), the Dirichlet 
conditions can be implemented by the insertion of the following 
 integrals representing $\delta$-functions\footnote{One might worry 
that imposition of the Dirichlet conditions on top of the Neumann 
conditions already built in is overconstraining. Classically, 
 it is a legitimate concern. But in the path integral 
calculation, the effect of the extra Neumann conditions is 
 to restrict the behavior of the functions only in the infinitesimal 
 vicinity of the ends of the interval and this modification is of 
 measure zero in the space of functions to be path-integrated.}:
\eqabegin
 \int d\mu_n \exp\left( {i\mu_n\over \sqrt{s}} (a_{n,0} + \sqrt{2} \sum_{m=1}^\infty a_{n,m})\right) \comma \label{sourcemu}\\
 \int d\mu'_n \exp\left( {i\mu'_n\over \sqrt{s}} (a_{n,0} + \sqrt{2} \sum_{m=1}^\infty a_{n,m}(-1)^m )\right) \period \label{sourcemup}
\eqaend
The terms in the exponent can then be regarded as an 
 addition of extra source terms  of the form 
\eqabegin
 \Delta j_{n,0} &=& {1\over a^2 \sqrt{s}}(\mu_n+\mu'_n) \comma \\
\Delta j_{n,m} &=& {\sqrt{2} \over a^2 \sqrt{s}} 
 (\mu_n+ (-1)^m \mu'_n) \comma 
\eqaend
which, in the light of definitions (\ref{jn0}), (\ref{jnm}), can be
written simply as 
\eqabegin
 \Delta j_n(\rho) &=& {2e^{-2\rho} \over R_1^2}(\mu_n \delta(\rho) 
 +\mu'_n\delta(s-\rho))\period
\eqaend
In other words, we are putting random sources at the ends to 
enforce the Dirichlet conditions. Replacing $j_n(\rho)$ in 
 (\ref{expj}) by $j_n(\rho)+\Delta j_n(\rho)$, we get, after 
 some calculations, the following additional terms for the exponent:
\eqabegin
 \Delta E(j)_n &=& = -4\pi\al'\left( N_n^0 \tilde{\mu}^2_n 
 + {1\over n^2 N_n^0}\tilde{\mu'}^2_n \right) \nn\\
&& \quad  +\pi\al'R_1^4 {N_n^0 \over \sinh^2 ns }
\left(  (\calJ_n(s))^2\tanh^2 ns + \left( \calJ_n(0) -{1\over \cosh ns}
\calJ_n(s)\right)^2\right)\comma \nn\\
&& \label{deltaexpj}
\eqaend
where 
\eqabegin
 N_n^0 &\equiv & N_n(0,0) = N_n(s,s)= 
{\cosh ns \over n\sinh ns }\comma \\
\calJ_n(s) &\equiv & \int_0^s d\rho e^{2\rho} j_n(\rho) \cosh n(s-\rho) 
\comma \label{caljns}\\
\calJ_n(0) &\equiv & \int_0^s d\rho e^{2\rho} j_n(\rho)\cosh n\rho  
\period \label{caljn0}
\eqaend
Modified sources   
 $\tilde{\mu}_n$ and $\tilde{\mu'}_n$   
 are introduced  in perfecting the squares but the Jacobian factor
 is unity  for this rewriting. Upon integration over 
 these sources, the determinant factors cancel, except for a factor 
 of $n$. Taking the product over $n$, we then get $n_0 (\prod_{n=1}
^\infty n)^2 \sim n_0$, up to a constant. This extra factor of $n_0$
 cancels $1/n_0$ in (\ref{dneu}) and we reproduce (\ref{ddir}).
On the other hand, the remaining terms in (\ref{deltaexpj}) give, 
after some rearrangements, 
\eqabegin
-\pi\al'R_1^4 \left( -\int_0^sd\rho\int_0^sd\rho'
e^{2(\rho+\rho')}j_n(\rho)j_n(\rho')
{2\over 2n\sinh ns}\cosh n(s-(\rho+\rho'))\right)
\comma 
\eqaend
which when added to the original piece (\ref{expj}) precisely 
effects the change  $N_n(\rho,\rho') \rightarrow D_n(\rho,\rho')$. 
This demonstrates that our extra-source method indeed works nicely. 
\newpage
\shead{$\xi_1$-$\xi_2$ sector}
With this warm up, let us now proceed to the main task of 
 incorporating the boundary conditions for the $\xi_1$-$\xi_2$ system. 
Since the basic method has already been described in detail,
 we shall only give the essence. 
 \par
The extra source terms for $x_{1,n}(\rho)$ and $x_{2,n}(\rho)$   
to be added now take the form 
\eqabegin
\Delta j^1_n(\rho) &=& {2e^{-2\rho}\over R_1^2}\mu'_n
 \delta(s-\rho)\sinh\chi
\comma \\
\Delta j^2_n(\rho) &=& {2e^{-2\rho}\over R_1^2}
(\mu \delta(\rho) +\mu'_n\delta(s-\rho)\cosh\chi ) \period
\eqaend
Substituting the shifted expressions $j^i_n(\rho) +\Delta j^i_n(\rho)$ into
 the exponent (\ref{expj}), where now we must replace $j_n(\rho)
N(\rho,\rho')j_n(\rho')$ by $j_n^i (\rho)\eta_{ij} N(\rho,\rho')j^j_n(\rho')$
 with the metric $\eta_{ij} =(-,+)$, we get  after some calculations
\eqabegin
E_n(j) &=& -4\pi\al' \left( N_n^0 \tilde{\mu}_n^2 + {N_n^0H_n \over 
\cosh^2 ns} \tilde{\mu'}_n^2\right) \nn\\
&& \quad + \pi \al'R_1^4 \Biggl[ {1\over n\sinh ns \cosh ns}\calJ^2_n(s)
\calJ^2_n(s) \nn\\
&& \quad + {\cosh ns \over n H_n \sinh ns } \left( \calJ^2_n(0)\cosh\chi
 -\calJ^1_n(0)\sinh\chi -{\cosh\chi \over \cosh ns}\calJ^2_n(s)\right)^2
\Biggr] \comma  \label{expj12}
\eqaend
where the quantity $H_n$ is defined by 
\eqabegin
H_n &\equiv & \sinh^2 ns -\sinh^2\chi ={1\over 4}e^{2ns}
(1-e^{2\chi -2ns})(1-e^{-2\chi-2ns}) \comma  \label{hn}
\eqaend
 and $\calJ_n^i(s), \calJ_n^i(0)$ are defined just as in (\ref{caljns}) 
 $\sim$ (\ref{caljn0}) with $j_n^i(\rho)$ in place of $j_n(\rho)$. 
Integrating over the sources $\tilde{\mu}_n, \tilde{\mu'}_n$, and forming  
 the appropriate products over   $n$ (with the contribution of the 
 primed system as well), the determinant factor  becomes 
\eqabegin
\tilde{D}^{(12)}(s) &\sim & {n_0^2 s\over \sinh\chi} \eta(is/\pi)^2 
 e^{2s/ 12} \prod_{n=1}^\infty (1-e^{2\chi -2ns})^{-1}
(1-e^{-2\chi-2ns})^{-1} \comma  \label{d12ap}
\eqaend
where the factor $n_0^2s/\sinh\chi$ is due to the $n=0$ mode. 
This must be multiplied by the already existing factor $D^{(N)}(s)^2
 \sim (1/n_0^2 s)\eta(is/\pi)^{-2}$  for two degrees of freedom. 
The factor $n_0^2s$ and the $\eta$-functions cancel and we get 
\eqabegin
D^{(12)}(s) &\sim &  {1\over \sinh\chi} 
 e^{2s/ 12} \prod_{n=1}^\infty (1-e^{2\chi -2ns})^{-1}
(1-e^{-2\chi-2ns})^{-1} \period \label{d12}
\eqaend
As for the Green's function,  we can read off from  
 (\ref{expj12}) the corrections $\Delta G_{ij,n}$  to be added to 
 $\eta_{ij} N_n(\rho,\rho')$.  Omitting the details, the final result for 
 the full Green's functions is 
\newpage
\eqabegin
G_{11,n}(\rho,\rho') &=& {1\over 2n}(\sinh n(\rho+\rho')
 +\sinh n|\rho-\rho'|)\nn\\
&& \quad  -{\sinh ns \cosh ns \over 2n H_n}
(\cosh n(\rho+\rho')+\cosh n(\rho-\rho'))    \comma \label{gn11}\\
G_{12,n}(\rho,\rho')&=& {\sinh 2\chi \over 4nH_n}(\sinh n(\rho+\rho') 
 -\sinh n(\rho -\rho')) \comma \label{gn12}\\
G_{21,n}(\rho,\rho')&=& {\sinh 2\chi \over 4nH_n}(\sinh n(\rho+\rho') 
 +\sinh n(\rho -\rho')) \comma\label{gn21} \\
G_{22,n}(\rho,\rho') &=&{1\over 2n}(\sinh n(\rho+\rho')
 -\sinh n|\rho-\rho'|)\nn\\
&& \quad  -{\sinh ns \cosh ns \over 2n H_n}
(\cosh n(\rho+\rho')-\cosh n(\rho-\rho'))      \period \label{gn22}
\eqaend
One can check that they satisfy $(-\del_\rho^2 +n^2)G_{ij,n}(\rho,\rho')
 =\eta_{ij} \delta(\rho-\rho')$ and that in the limit $\chi \rightarrow 0$ 
  they reduce to the expected forms, namely,  
 $G_{11,n} \rightarrow -N_n$, $G_{22,n} \rightarrow D_n$, and 
$G_{12,n}, G_{21,n} \rightarrow 0$. Moreover, 
 they are of  such  forms  that we can take the limit $n\rightarrow 0$ to 
 obtain  the Green's functions for the zero mode as well. \par
%
\shead{Summary of the lowest order amplitude}
Let us now assemble the results so far obtained and write down 
 the $0$-th-order amplitude, \ie  yet without  the corrections at the 
boundaries.  It consists of  the contribution from 
  the classical part  $e^{-S_{cl}}$ with $S_{cl}$ given in (\ref{scl}), 
 the one  from  
 each transverse coordinate 
 $\xi_I$  given in (\ref{ddir}), and  the one due to  the  $\xi_1$-$\xi_2$
 system,   (\ref{d12}). To these we must add the contribution 
of the $b$-$c$ ghosts, which is , as usual, $D^{(gh)}(s) = \eta(is/\pi)^2$.
Finally, we must integrate over the proper time $s$. Since we are 
 dealing with  a cylinder, the proper measure should simply be $\int_0^\infty
 ds$.  Thus, using the explicit form of $\eta(is/\pi)$, we get  
\eqabegin
\calV_0(f_1,f_2)  &\sim &  \int_0^\infty ds e^{2s} e^{(D-26)s/12} s^{-(D-2)/2} e^{-(f_2-f_1)_\perp^2/2\al's }\nn\\
&& \quad \cdot {1\over \sinh\chi}
 \prod_{n=1}^\infty  \left( 1-e^{-2ns}\right) ^{-(D-4)}
 \left( 1-e^{-2ns+2\chi}\right)^{-1}
 \left(1-e^{-2ns-2\chi}\right)^{-1}  \label{loamp}
\eqaend
as our lowest order approximation to the amplitude with 
 fixed trajectories.  In terms of the usual $\theta$- and $\eta$-functions, it can  be written compactly as 
\eqabegin
\calV_0(f_1,f_2)  &\sim &  \int_0^\infty ds s^{-(D-2)/2} e^{-(f_2-f_1)_\perp^2/2\al's } \eta(\tau)^{-(D-2)} {\theta'_1(0|\tau)
 \over \theta_1(\nu|\tau)} \comma \label{loampth}
\eqaend
where $\tau =is/\pi$ and $\nu =-i\chi/\pi$. 
\subsection{Corrections at the boundaries}
We shall now compute the corrections due to the extendedness of the 
 boundaries, \ie to the fluctuations of the geodesic coordinates 
$\zeta_i(\theta)$. \par
The first task is to derive the effective boundary action for 
$\zeta_i(\theta)$ arising from the integration over the special sources 
 $\nu^\mu(R_i,\theta)$ that we introduced in (\ref{scst}) to enforce 
 the constraints (\ref{cstxitil}). These sources have until now been 
 represented by the general source $J^A(z)$ (or its 
 Fourier components $j^A_n$)  and the integration over
 the string coordinates $\xi^\mu(z)$ produced an exponential 
 factor with the exponent quadratic in $J^A_n$'s connected by 
 the Green's functions. \par
To apply this to the present case, we must 
recall the following: (i) $\nu^\mu(R_i,\theta)$'s do not have 
 constant parts, since they are designed to couple only to the 
 non-constant part of $\xi_\mu$. 
(ii) Because, at each boundary, components of $\xi_\mu$ orthogonal 
 to the direction of the trajectory vanish by the boundary conditions, 
 the only components of $\nu^\mu(R_i,\theta)$ that actually 
 couple to $\xi_\mu$ are the ones along the trajectory, namely
$\nu(R_1,\theta)\cdot u_1$ and $\nu(R_2, \theta)\cdot u_2$. 
The point (i) says that we only need a Green's function in the space 
 of non-zero modes (with respect to $\theta$). Furthermore, 
 the point (ii) tells us that only the Green's functions with components
 in the trajectory plane will be required. \par
Thus we form the following two-dimensional Green's function on the 
 worldsheet without the zero mode:
\eqabegin
 G_{ij}(z,z')=G_{ij}(\rho,\theta;\rho',\theta')  &\equiv & 
\sum_{n\ge 1}G_{ij,n}(\rho,\rho) \cos n(\theta -\theta') \comma 
\label{2dgrfn}
\eqaend
where $G_{ij,n}(\rho,\rho')$'s are as given in (\ref{gn11})
 $\sim$ (\ref{gn22}).
 This 
satisfies $-\del_z^2 G_{ij}(z,z') = \pi \tilde{\delta}^2(z,z')$, 
 where $\tilde{\delta}^2(z,z')\equiv (1/R_1^2)e^{-2\rho}
\delta(\rho-\rho')(\delta(\theta -\theta') -(1/2\pi))$ is the 
 delta function in the space of non-zero modes. Then defining 
\eqabegin
\nu^i(\theta) &\equiv & \nu(R_i, \theta)\cdot u^i \comma 
\eqaend
one finds that 
the exponential factor quadratic in the source takes the form  
\eqabegin
\exp\left( -\al' \int d\rho d\rho' d\theta d\theta' j^i(\rho,\theta)
G_{ij}(\rho,\theta;\rho',\theta')j^j(\rho',\theta')\right) \comma 
\eqaend
where 
\eqabegin
j^1(\rho,\theta) &=& \nu^1(\theta)\delta(\rho) + \nu^2(\theta)
 \delta(s-\rho)\cosh\chi \comma \\
j^2(\rho,\theta) &=& \nu^2(\theta) \delta(s-\rho)\sinh\chi\period
\eqaend
Using the explicit forms of $G_{ij,n}(\rho,\rho')$, this exponential  
 can be written as 
\eqabegin 
\exp\left( -\al' \int d\theta d\theta' \nu^i(\theta) 
\calG_{ij}(\theta,\theta') \nu^j(\theta')\right)\comma 
\eqaend
with 
\eqabegin
\calG_{ij}(\theta,\theta')  &=& -\sum_{n=1}^\infty {\sinh ns \over nH_n}\matrixii{\cosh ns}{\cosh\chi}{
\cosh\chi}{\cosh ns}  _{ij} \cos n(\theta -\theta')\period\label{calgij}
\eqaend
Note that at this stage the Green's function,  
 $\calG_{ij}(\theta,\theta')$,  has become completely symmetric in $(i,j)$.  
This must be the case because  the indices  $i=1,2 $  for   $\nu^i(\theta)$
 now refer to the two trajectories symmetrically.  
\par
We must further add to the exponent  the  remaining terms linear in the 
 sources that directly couple to $\zeta_i(\theta)$'s (see (\ref{scst})). 
 It is convenient  to separate  out the parts containing $\nu^i(\theta)$ and 
 write them  in the form 
\eqabegin
&& i\int d\theta \left( \sum_i \nu^i(\theta) Z_i(\theta)
 -\Omega(\theta) \right) \comma \label{linnu}
\eqaend
where 
\eqabegin
Z_i(\theta)&\equiv & \sqrt{-h_i}\, \zeta_i(\theta) 
 +{1\over 3!}{K_i^2 \over \sqrt{-h_i}}\widetilde{ \zeta_i(\theta)^3}
  + \cdots \comma  \\
 \Omega(\theta) &\equiv & \sum_i \nu_\mu(R_i,\theta) P_i^{\mu\nu}
\left( \half K_{i,\nu} \widetilde{\zeta_i(\theta)^2} + \cdots\right)\period \\
\eqaend
In the above,   $\widetilde{\zeta_i(\theta)^n}$  means that  the 
 constant part  contained in  $\zeta_i(\theta)^n$ is removed.  
 Then integration over $\nu^i(\theta)$ 
 yields 
\eqabegin
(\det \calG)^{-1/2} \exp\left( {1\over 4\al'}\int d\theta d\theta' Z_i(\theta) \calG^{-1}(\theta,\theta')^{ij} Z_j(\theta') \right) \comma \label{zgz}
\eqaend
where the inverse Green's function $\calG^{-1}(\theta,\theta')$ is 
 easily obtained from (\ref{calgij}) as  
\eqabegin
\calG^{-1}(\theta,\theta') &=& -{1\over \pi^2} 
\sum_{n=1}^\infty {\sinh ns \over nH_n}\matrixii{\cosh ns}{-\cosh\chi}{
-\cosh\chi}{\cosh ns}   \cos n(\theta -\theta')\period\label{calgijinv}
\eqaend
\par
We have now completed the calculation of  the effective interaction 
 of $\zeta_i(\theta)$'s. It consists of the factors $e^{-S_{cl-q}}$ 
 from (\ref{sclq}), $e^{-i\int d\theta \Omega(\theta)}$ from (\ref{linnu}), 
 and the factor (\ref{zgz}) just computed.  Note that  {\it so far  
 no approximation has been made.}
\par
To proceed further, we now need to make an approximation. As was 
 already announced, we will consider the case where the accelerations 
  $\ddot{f}_i$ ( and higher $t_i$-derivatives of $f_i^\mu(t_i)$ ) are small
 and treat the terms containing them as perturbations.  
 Specifically, in this article we will keep only the terms linear in  the 
 extrinsic curvature $K^\mu =P^{\mu\nu} \ddot{f}_\nu$. Then, 
 the expression we need to  deal with is 
\newpage
\eqabegin
\calV_{corr} &=& (\det \calG)^{-1/2} \exp\left( 
{1\over 4\al'}\int d\theta d\theta'  \sqrt{-h_i}\, \zeta_i(\theta) \calG^{-1}(\theta,\theta')^{ij}
 \sqrt{-h_j}\, \zeta_j(\theta') \right)\nn\\
&& \quad \times \Biggl[ 1-{1\over 2\al' s}(f_2-f_1)_{\perp\mu}
 \int {d\theta\over 2\pi} \left(K_2^\mu\zeta_2(\theta)^2 
-K_1^\mu\zeta_1(\theta)^2\right)\nn\\
&& \quad + \half \sum_i \int d\theta \nu_\mu(R_i,\theta)
P^{\mu\nu}_i K_{i,\nu} \widetilde{\zeta_i(\theta)^2} \Biggr]\period
\label{vcorr}
\eqaend
The Gaussian integration over $\zeta_i(\theta)$ yields the following 
 effects.  First, it produces the determinant factor, which consists of  
$(\det \calG)^{1/2}$ which cancels $ (\det \calG)^{-1/2}$ in front  and 
\eqabegin
 (\prod_{n=1}(-h_1/\al')^{-1/2})^2  (\prod_{n=1}(-h_2/\al')^{-1/2})^2
&=& \sqrt{-h_1/\al'}\, \sqrt{-h_2/\al'}\comma 
\eqaend
which is obtained by the Fourier mode expansion and the $\zeta$-function
 regularization. The factor $\sqrt{-h_i}$ will be  important in changing $dt_i$ 
  into  the  proper time $d\tau_i\sim dt \sqrt{-h_i}$.   
 Second, the terms quadratic in $\zeta_i(\theta)$ are 
 replaced by their quantum averages.   The relevant  
 correlation function $\langle \zeta_i(\theta) \zeta_j(\theta')\rangle$  
 is given by  
\eqabegin
\langle \zeta_i(\theta) \zeta_j(\theta')\rangle
&=& -{2\al'\over \sqrt{-h_i}\sqrt{-h_j} }\calG_{ij}(\theta,\theta')\period
\label{gfzeta}
\eqaend
In particular, using the explicit form of the Green's function, 
 the ones  at the coincident point  are computed as  
\eqabegin
\langle \zeta_i(\theta)^2\rangle &=& {2\al'\over h_i } 
\calG_{ii}(\theta,\theta)
= {2\al'\over h_i  }\left( \ln \ep -\sum_{n=1}^\infty {1\over nH_n}
(\cosh2\chi -e^{-2ns})\right)\comma \label{zetasq}
\eqaend
where we have regularized the divergent sum 
  $\sum_{n\ge 1}(1/n) \rightarrow 
 \sum_{n\ge 1} (e^{-\ep n}/n) =-\ln \ep $. Since this is independent of 
 $\theta$, we immediately have $\langle \widetilde{\zeta_i(\theta)^2}
\rangle=0$, so that the last term in (\ref{vcorr}) does not contribute 
 to this order.  As for the term containing $f_2-f_1$, we get a 
 divergent term as well as a fininte correction. The former is of the form 
\eqabegin
  -{1\over s}(f_2-f_1)_\perp\cdot \left({K_2 \over h_2} 
-{K_1\over h_1}\right) \ln \ep \period \label{lnep}
\eqaend
If we remember that we have,   in the lowest order 
 amplitude,  the classical part of the 
 form  $\exp(-(1/2\al' s)(f_2-f_1)_\perp^2)$, we see that (\ref{lnep})
 can be  absorbed by the renormalization of the trajectories (subscript
 $R$ stands for renormalized  quantities)  
\eqabegin
f_i^\mu &=& f_{i,R}^\mu +\delta f_{i,R}^\mu \comma \\
\delta f_{i,R}^\mu&=& -\al' {K^\mu_{i,R} \over h_i} \ln \ep \period \label{renf}
\eqaend
 It is gratifying that  (\ref{renf})  is exactly   the form 
 found for the one D-particle case treated in \cite{paperI}\footnote{
We have also  confirmed that divergences that occur at 
  the second order in $K^\mu$ are  precisely absorbed by the 
 renormalization of $\sqrt{-h_1}\,\sqrt{-h_2}$ with the same 
 prescription (\ref{renf}). This is a highly non-trivial check of the 
 consistency of our treatment.}. \par
Putting all the findings together, we now have the final form of the 
 amplitude for fixed trajectories, valid up to the first order in $\ddot{f}_i$:
\eqabegin
\calV(f_1,f_2) &=&  \int \prod_i dt_i\sqrt{-h_i/\al'} 
  \int_0^\infty ds e^{2s} e^{(D-26)s/12} s^{-(D-2/2)} e^{-(f_2-f_1)_\perp^2/2\al's }\nn\\
&& \quad \cdot {1\over \sinh\chi}
 \prod_{n=1}^\infty  \left( 1-e^{-2ns}\right) ^{-(D-4)}
 \left( 1-e^{-2ns+2\chi}\right)^{-1}
 \left(1-e^{-2ns-2\chi}\right)^{-1}  \nn\\
&&  \quad \cdot e^{(f_2-f_1)_\perp \cdot \Delta K} \comma \label{fixedamp}\\
\where \Delta K^\mu &\equiv & {1\over s} \left({K^\mu_2 \over h_2} 
-{K^\mu_1\over h_1}\right)\sum_{n=1}^\infty {1\over nH_n}
(\cosh2\chi -e^{-2ns}) \period
\eqaend
\subsection{Amplitude for quantized D-particles}
\shead{Quantization of the collective coordinates}
Having obtained the amplitude for fixed trajectories, we  
 now quantize the trajectories themselves. 
The basic formalism developed  in \cite{paperI} works with a slight 
 modification and hence we shall  only describe the essence
 and refer the reader  to \cite{paperI}  for further techinical  details. 
 \par 
 To the accuracy of our approximation, the action for a 
 relativistic D-particle is of the form 
$S_0 =-im_0 \int_0^1 dt \sqrt{-\fdot^2})$ (with $m_0 $  the bare mass), 
 which,  following \cite{Polyakov}, 
 can be effectively turned into a more tractable  quadratic action
\eqabegin
 S &=& \int_0^T d\tau \half (\fdot^2(\tau) -m^2) \period \label{dpaction} 
\eqaend
Here,  $m$ is the renormalized mass and 
  $\tau(t)=(1/m) \int_0^tdt' \sqrt{-h(t')} $
  is  the  (rescaled) proper time. Then the $t$-derivatives can be 
 rewritten into  $\tau$-derivatives  in the manner 
 $\fdot(t)/\sqrt{-h(t)}$ $ = \fdot (\tau)/m$, $\ddot{f} (t)/(-h(t))
 = \ddot{f}(\tau)/m^2$ , etc. Hereafter, dots refer to $\tau$-derivatives.  
 The quantum amplitude is then obtained 
 by  $\calA = \int_0^\infty dT \int \calD f(\tau)\exp(iS) 
\calV\left(f\right)$, where $\calV\left(f\right)$ is the amplitude 
( or the \lq\lq vertex operator" rather) for fixed trajectory. \par
For the present case, $\calV\left(f\right)$ computed in the 
 previous subsection depends on $\fdot_i$ and $\ddot{f}_i$ as well as 
 on $f_i$. To separate out the dependence on the derivatives, it 
 is convenient to replace them by  new variables 
 $v_{n,i}$,   where $n (=1,2)$ refers to the number of derivatives,  
  via the insertion of unity:
\eqabegin
1 &=& \int \prod_{n,i} dv_{n,i}   \int \prod_{n,i}d\omega_{n,i}
 \exp\left( i\sum_{n,i} \omega_{n,i} (v_{n,i} -\del^n f_i (\tau_i)) 
 \right)\period \label{insunity}
\eqaend
 Then  the path integral over $f_i(\tau_i)$ we need to consider 
 at the first stage is 
\eqabegin
\calA_f &\equiv & \int \prod_i \calD f_i(\tau_i)  \exp\left( i \sum_i 
\half \int_0^{T_i} d\tau   ( \fdot_i^2(\tau) -m^2) \right) \nn\\
&& \cdot \int d\tau_1 d\tau_2 \exp\left( -(f_2(\tau_2) 
-f_1(\tau_1))_\perp^2/(2\al' s) + (f_2-f_1)_\perp \cdot \Delta K(v_{n,i}) 
  \right)\nn\\
&& \cdot  \exp\left(- i\sum_{n,i} \omega_{n,i} \cdot \del^n
 f_i (\tau_i))
 \right)\period \label{ampf}
\eqaend
Let us denote by $f_i$ and $f'_i$  the initial and the final 
 positions of the D-particles respectively, and  
decompose $f^\mu_i(\tau_i)$ into the classical part $f_{cl,i}^\mu (\tau_i)$
 and the quantum 
part $\ftil_i^\mu(\tau_i)$:
\eqabegin
 f_i^\mu(\tau_i) &=& f^\mu_{cl, i}(\tau_i) + \ftil^\mu_i(\tau_i) 
\comma \\
f^\mu_{cl, i}(\tau_i) &=& {y_i^\mu \over T_i} \tau_i + f^\mu_i 
\comma \\
\where y_i & \equiv & f'_i -f_i \comma \\
 \ftil_i(0) &=& \ftil_i(T_i) =0 \period \label{bcftil} 
\eqaend
Then due to the boundary condition (\ref{bcftil}) the classical and 
 the quantum parts separate in the action:
\eqabegin
\half  \int_0^{T_i} d\tau (\fdot_i^2-m^2) 
&=& {y_i^2 \over 2T_i} -\half m^2 T_i + \half  \int_0^{T_i} d\tau 
{\dot{\ftil}_i}^2 \period
\eqaend
On the other hand,  for the term $(f_2(\tau_2)-f_1(\tau_1))_\perp^2$ 
 such a separation appears difficult at first sight.   
 This problem,however, is  solved by going to the  momentum representation. \par
Let $p_i$ and $p'_i$ be the incoming and the outgoing momenta 
 of the D-particles respectively  as in Fig.3. 
\begin{center} 
\epsfxsize=4cm
\quad\epsfbox{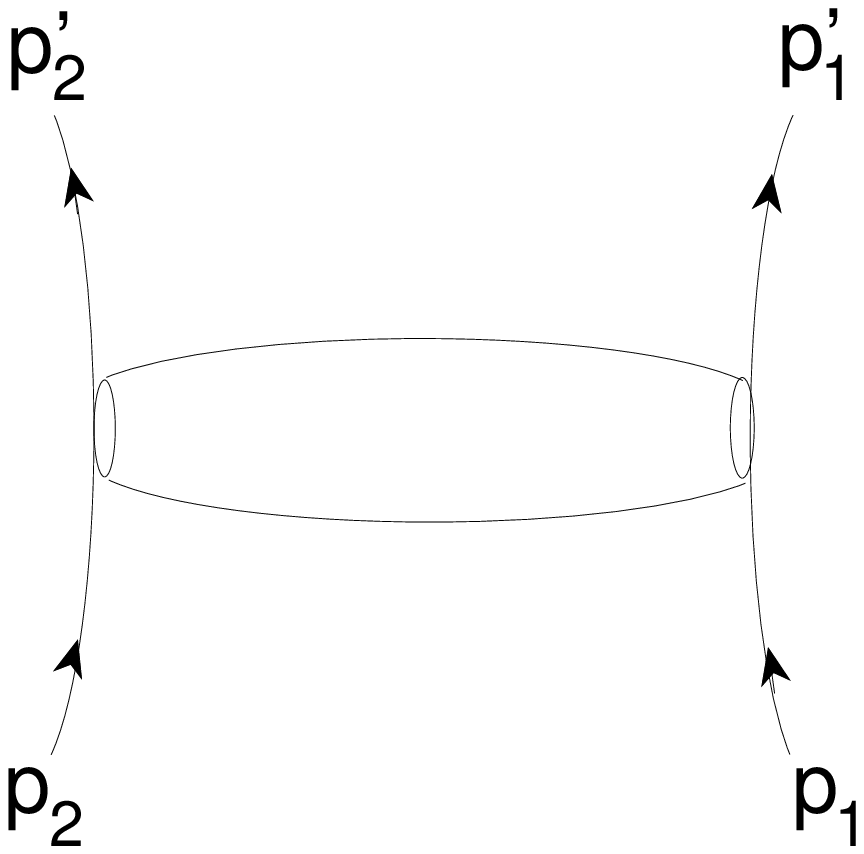} \\
{\small {\bf Fig.3} \quad Scattering diagram in momentum space.}
\end{center}
\parn
%
%
%
Then the factor
 to be multiplied for going to the momentum representation is 
\eqabegin
&&  \exp(ip_1\cdot f_1 -ip'_1\cdot f'_1)
 \exp(ip_2\cdot f_2 -ip'_2\cdot f'_2) \nn\\
&&\quad = \exp(-i(y_1\cdot p'_1+y_2\cdot p'_2))
 \exp(i(f\cdot(p_1-p'_1)
 + f_2\cdot (p_1+p_2 -p'_1 -p'_2))) \comma \nn\\
\eqaend
where we have defined $f\equiv f_1-f_2$. Integration over $f_2$ 
 is now trivial and produces the momentum conserving 
$\delta$-function  $\delta (p_1+p_2 -p'_1 -p'_2)$. As for 
 the integration over $f$,  
it  is of the form 
\eqabegin
I_f &=&  \int d^Df e^{if\cdot(p_1-p'_1)} e^{-(f+g)_\perp^2/(2\al' s)
 +(f+g)_\perp \cdot \Delta K} \comma \\
\where g &\equiv & {y_1 \over T_1}\tau_1 -{y_2 \over T_2}\tau_2
 + \ftil_1(\tau_1) -\ftil_2(\tau_2)\period
\eqaend
So by shifting $f+g \rightarrow f$, we get
\eqabegin
 I_f &=&  e^{-ig\cdot(p_1-p'_1)} \int d^Df 
 e^{if\cdot(p_1-p'_1)}  e^{-f_\perp^2/(2\al' s)
 + f_\perp\cdot \Delta K} \nn\\
 &\sim&  s^{(D-2)/2}\,  e^{ -ig\cdot(p_1-p'_1)} e^{-(\al's/2)
 (p_1-p'_1 -i\Delta K)_\perp^2} \delta^2((p'_1-p_1)_\|)\period
 \label{intf}
\eqaend
 $\delta^2((p'_1-p_1)_\|)$ is the two-dimensional $\delta$-function
 for the components in the plane spanned by the vectors $v_{i,1}$'s. 
Note that since $g$ in the exponent is linear in $y_i$, the 
 classical and the quantum parts are now separated. 
 Note also that the front factor  cancels  
 $s^{-(D-2)/2}$ already present in the amplitude.  \par
Next we perform the integration over $y_i$. Gathering together 
 the terms containing $y_i$, the integral is 
\eqabegin
 I_y &=& \int dy_1 dy_2 
\exp\left\{ i \sum_i \left( {1\over 2T_i}y_i^2 
 -{\omega_{1,i}+(p_i -p'_i)\tau_i \over T_i}\cdot y_i
\right) \right\} \nn\\
&\sim & (T_1T_2)^{D/2} 
\exp\left( -\sum_i{i\over 2T_i} \left( \omega_{1,i}+p'_i T_i + 
 \tau_i (p_i-p'_i)\right)^2\right) \comma \label{inty}
\eqaend
where we have used  momentum conservation. \par
Next we deal with  the quantum fluctuations
 $\ftil_i$. The integral to be performed is 
\eqabegin
 I_\ftil &=& \prod_{i=1}^2 \int_{\ftil_i(0)=\ftil_i(T)=0}\calD \ftil_i 
 \exp \left( 
 {i\over 2} \int_0^T d\tau \dot{\ftil}_i^2(\tau)  
-\sum_n i\omega_{n,i}\cdot 
\del^nf_i(\tau_i)+i(p'_i-p_i)\cdot\ftil_i(\tau_i) 
 \right) \period \nn
\\
\eqaend
This is precisely of the type that occured in 
\cite{paperI} and can easily be evaluated by using the one-dimensional 
Green's function satisfying the Dirichlet conditions at the ends of 
 the interval $\left[0,T\right]$. As it was shown there, 
 with the standard $\zeta$-function regularization, 
 the contributions of $\del^n\ftil(\tau_i)$ average out to zero 
 for $n\ge 2$. Skipping the details, the result is (using again 
 the momentum conservation) 
\eqabegin
I_\ftil &=& (T_1T_2)^{-D/2} \exp\left( \sum_i 
i{\omega_{1,i}^2 \over 2T_i} -i{(p_i'-p_i)\cdot
 \omega_{1,i} \over T_i} \left( \tau_i
 -{T_i\over 2}\right) +{i \over 2T_i} (p'_i-p_i)^2 
\tau_i(\tau_i -T_i)\right) \period \nn\\
\label{intftil}
\eqaend
Combining (\ref{inty}) and (\ref{intftil}), the front factors cancel 
 and the exponent becomes 
\eqabegin
 \sum_i \left( -{i\over 2}\omega_{1,i}  (p_i + p'_i)
 + {i\over 2}({p'}^2_i -p_i^2)\tau_i -{i\over 2}{p_i'}^2 T_i
\right) \period
\eqaend
If we now put back the factors $\exp( i\sum_{n,i} \omega_{n,i}v_{n,i})$ 
  (see (\ref{insunity})) 
and integrate over $\omega_{n,i}$, we get the following 
 $\delta$-functions:
\eqabegin
&& \prod_i \delta(v_{1,i} -\half ( p_i + p'_i)) \delta (v_{2,i}) 
\period
\eqaend
This means that effectively we can replace $\fdot_i$ 
 by $\half ( p_i + p'_i)$ and set $\ddot{f}_i$ to zero, exactly 
 the same rule that we established in \cite{paperI}. Therefore 
 the correction $\Delta K$ that remained in (\ref{intf}) 
 actually vanishes. Another consequence is that 
 what has been referred to as the \lq\lq trajectory plane" should now 
 be understood as the one spanned by the mean momenta 
 $\half(p_1'+p_1)$ and $\half(p_2'+p_2)$. Then when the D-particles 
 are put on shell, we easily see that the momentum transfer 
 $k=p'_1 -p_1 = -(p'_2-p_2)$ is orthogonal to this plane since 
\eqabegin
 k \cdot \half (p'_1 +p_1) &=& \half({p'}_1^2 + m^2) 
 -\half(p_1^2 + m^2) =0 \comma \\
 -k\cdot \half (p'_2 +p_2) &=& \half({p'}_2^2 + m^2) 
 -\half(p_2^2 + m^2) =0 \period
\eqaend
This is quite consistent with the presence of $\delta^2((p'_1-p_1)_\|)$
 in (\ref{intf}) and allows us to replace $(p'_1-p_1)_\perp^2$ in 
 the exponent  by   $(p'_1-p_1)^2$.  
\par
The remaining integrations over the interaction points $\tau_i$  and 
 over $T_i$ are trivial and they produce the  four propagator 
 legs\cite{paperI}, to be removed from the proper amplitude. 
\par
We may now write down the quantized proper amplitude in complete form. 
Omitting the momentum conserving $\delta$-function and with 
 the understanding that the D-particles are put on shell, it reads 
\eqabegin
\calA_{bos}(p_1,p_2; p'_1,p'_2) &\sim &
  \int_0^\infty ds e^{2s} e^{(D-26)s/12} 
e^{-(\al's/2)(p'_1-p_1)^2}\nn\\
&& \quad \cdot {1\over \sinh\chi}
 \prod_{n=1}^\infty  \left( 1-e^{-2ns}\right) ^{-(D-4)}
 \left( 1-e^{-2ns+2\chi}\right)^{-1}
 \left(1-e^{-2ns-2\chi}\right)^{-1} \comma\nn\\
 \label{fullamp}
\eqaend
where $\chi$ is defined by 
\eqabegin
\cosh\chi &=& -u_1\cdot u_2 \comma \label{coshchi} \\
u_i &=& {p_i+p'_i \over \sqrt{-(p_i+p'_i)^2}} \period \label{ui}
\eqaend
This, however, is not quite the correct answer for fully quantum 
 amplitude: We have not yet taken into account the quantum 
 indistinguishability of D-particles. It is clearly seen 
 when we express $\cosh\chi$ defined above in terms 
 of the velocity $v$ and the scattering angle 
 $\theta_{CM}$ in the center of mass frame of the D-particles. 
It takes the form 
\eqabegin
 \cosh\chi &=& {1+v^2\cos^2(\theta_{CM}/2) \over 
1-v^2\cos^2(\theta_{CM}/2)} \period
\eqaend
One sees that , for any $v$, $\chi$ vanishes for $\theta_{CM} =\pi$, 
 \ie for back-scattering but it does not vanish for 
 the forward scattering. For indistinguishable particles this is 
 obviously incorrect. The cure of course is to add the amplitude 
 in which  $\theta_{CM}$ is replaced by $\pi -\theta_{CM}$. 
This replacement must also be done for the exponential factor 
$\exp\left(-(\al's/2)(p'_1-p_1)^2\right)$, 
which in the center of mass frame takes the form 
$\exp\left(-2\al' p^2 \sin^2(\theta_{CM}/2)\right)$, with $p$ the 
 maginitude of the spatial momentum. More generally the replacement 
 to be made is $p'_1 \leftrightarrow p'_2$, so that the desired 
 quantum amplitude is 
\eqabegin
\calA_{bos}(p_i,p'_i) &=& \calA_{bos}(p_1,p_2; p'_1,p'_2)
 + \calA_{bos}(p_1,p_2; p'_2,p'_1) \period
\eqaend
%
\shead{Forward scattering in the infinite mass limit}
Let us check our result against the known 
expression\cite{Bachas,BDC} 
 in the special case of forward scattering for infinitely heavy 
 D-particles.  In this case, the particles are treated as distinguishable 
 backgrounds and hence we should use (\ref{fullamp}) $\sim$ 
 (\ref{ui}). \par
To make the comparison, we first turn the amplitude into  
 the impact parameter representation by introducing  
  a transverse vector  $b^\mu$  and performing a Fourier transformation
 with respect to the momentum transfer $k_\perp =(p'_1-p_1)_\perp$. 
Since $\chi$ does not depend on the transverse components, this 
 transformation simply produces the factor 
\eqabegin
\int d^{D-2}k_\perp e^{ik_\perp \cdot b} e^{-(\al's/2) k_\perp^2} 
&\sim & s^{(D-2)/2} e^{-b^2/(2\al's)} \period \label{impactp}  
\eqaend
Let us now   
assume that the D-particles are moving in a common direction, 
 say 1,  with 
 velocities $V_1$ and $V_2$. Then the 4-momenta of the particles 
 are given by $p^\mu _i  = (E_i, E_iV_i,\vec{0}) $, where $E_i$ is the 
 energy of the particle $\# i$, and they do not change as they scatter.  
Hence, the unit-normalized vectors $u_i$ take a simple form 
\eqabegin
 u_i &=& {(1,V_i, \vec{0}) \over \sqrt{1-V_i^2}} \period 
\eqaend
From this $\sinh\chi$ is easily computed to be 
\eqabegin
\sinh\chi &=& \sqrt{(u_1\cdot u_2)^2 -1} = {|V_1-V_2| 
 \over \sqrt{1-V_1^2}\sqrt{1-V_2^2}} \comma \label{sinhchi}
\eqaend
which is exactly the expression that appears  in \cite{Bachas,BDC}. 
Applying (\ref{impactp}) and (\ref{sinhchi}) to (\ref{fullamp}), 
 we recover the known result in this special case.  
\section{Boundary State Approach}
\setcounter{equation}{0}
In this section, we shall demonstrate that the lowest order amplitude for 
fixed trajectories obtained in (\ref{loamp}) can be  
 reproduced  by using the boundary state representation 
 developed in \cite{Ishibashi,paperI}. 
 To avoid possible confusion, we wish to emphasize at  the outset that 
 the \lq\lq boundary state" that appears  below {\it does not} 
 correspond  to  a  definite state of a D-particle.  Rather it should 
 properly be understood as a representation of the D0-D0-string 
 interaction vertex describing the {\it transition} of a state of a D-particle
 into another state\footnote{This feature is obscured in the case of 
 infinitely heavy D-particles, since  then   their states do not change 
 by the interaction and the \lq\lq vertex" looks like  
 an ordinary state.}. This feature explicitly appears as the presence   
 of  the derivative $\fdot_i$, which, when  the D-particles are quantized
 as in the previous section, turns  into $\half (p_i+p'_i)$ containing both the initial and the final  momentum.    
\subsection{Boundary state representation of the vertex}
The boundary ket $\ket{B;f}$ associated with a trajectory $f^\mu(t)$ 
is of the structure 
\eqabegin
\ket{B;f}&=&\ket{B_z;f}\otimes 
\ket{B_{nz};f}\otimes\ket{B_{gh}}\comma 
\eqaend
where the subscripts \lq\lq z",
\lq\lq nz" and \lq\lq gh" stand for zero mode, non-zero modes and 
 ghosts, respectively. The explicit forms for the non-zero modes
 and the ghost parts for our problem are given by 
\eqabegin
\ket{B_{nz}; f_1} &=& \exp\left( -\sum_{n\ge1} {1\over n} 
 \al^\mu_{-n} D^\nu_\mu(\fdot_1) \altil_{-n, \nu}\right) \ket{0}\comma \\
\bra{B_{nz};f_2} &=& \bra{0} \exp\left( -\sum_{n\ge1} {1\over n} 
 \al^\mu_n D^\nu_\mu(\fdot_2) \altil_{n, \nu}\right) \comma \\
\ket{B_{gh}} &=&  \exp\left( \sum_{n=1}^\infty  
 \left[ \tilde{c}_{-n}b_{-n} + c_{-n}\tilde{b}_{-n}\right] \right) 
 (c_0+\tilde{c}_0) \ket{\downarrow\downarrow} \comma \\
\bra{B_{gh}} &=& \bra{\uparrow\uparrow} (b_0-\tilde{b}_0) 
 \exp\left( \sum_{n=1}^\infty  
 \left[ \tilde{c}_{n}b_{n} + c_{n}\tilde{b}_{n}\right] \right)\comma \\
\where D^\nu_\mu(\fdot_i) &=& (h_i)_\mu^\nu -(P_i)_\mu^\nu \period
\eqaend
The matrix $D^\nu_\mu(\fdot_i)$ has eigenvalue $+1$ along $\fdot_i$ 
(Neumann) direction and $-1$ for the transverse (Dirichlet) directions.
\par
As for the zero-mode part $\ket{B_z;f}$, it should essentially be the 
 position eigenstate $\ket{f}$ since the trajectory is given fixed. 
However, as was discussed in section 2.1, we must also 
 incorporate the requirement that the momentum transfer in the 
 tangential direction should vanish. This turned out to follow 
 automatically  from the requirement of BRST invariance. 
This invariance demands\cite{CallanII} 
$L_n \ket{B_z;f}\otimes \ket{B_{nz};f} = \tilde{L}_{-n}
\ket{B_z;f}\otimes \ket{B_{nz};f}$, where $L_n$ and $\tilde{L}_n$
 are the closed string Virasoro generators in the usual notation. 
Then, since $\al^\mu_n\ket{B_{nz};f} = -D^\mu_\nu(\fdot)\altil^\nu_{-n}
\ket{B_{nz};f}$ holds, one easily finds that we must have 
 $p^\mu D_\mu^\nu(\fdot)= -p^\nu$, or equivalently, 
$p^\mu u_\mu=0$  on $\ket{B_z;f}$, with $u^\mu$  the unit vector 
 along $\fdot^\mu$.  Thus the correct form of the 
 zero-mode part should be, up to a constant, 
\eqabegin
\ket{B_z;f_1} &=& \delta(p\cdot u_1)\ket{f_1} \comma\label{zmbket} \\
\bra{B_z;f_2} &=& \bra{f_2}\delta(p\cdot u_2) \period
\eqaend
One can check that in the case of infinitely heavy D-particles 
 moving parallel to each other  the ket (\ref{zmbket})  reduces to the one 
 constructed in \cite{BDC}. 
\subsection{Calculation of the amplitude}
The amplitude is now given by 
\eqabegin
 \calV_0(f_1, f_2) &=&  \bra{B;f_2} {1\over L_0+\tilde{L}_0 -2}
 U_0\ket{B; f_1}  \nn\\
&=& \int_0^\infty ds \bra{B;f_2} e^{-s (L_0+\tilde{L}_0 -2)} U_0
 \ket{B; f_1} \comma 
\eqaend
where 
\eqabegin
L_0+\tilde{L}_0-2&=& {\al'\over 2}p^2 +\sum_{n=1}^\infty 
 (\al_{-n}\cdot \al_n + \altil_{-n}\cdot \altil_n )\nn\\
&& \qquad  + \sum_n n:( b_{-n}c_n +\tilde{b}_{-n}\tilde{c}_n): -2 
\eqaend
is the usual closed string Hamiltonian and $U_0$ is 
 a ghost zero mode insertion, to be discussed below. 
\par
Let us now compute the  contribution to the amplitude from each sector. 
First, for the zero mode sector, we have, ignoring the overall constant,
\eqabegin
\calV_z&=& \bra{f_2}\delta(p\cdot u_2)
 e^{(-\al' s/2)p^2}\delta(p\cdot u_1) \ket{f_1} \nn\\
& \sim & \int d^Dp  e^{ip\cdot (f_2-f_1) 
 + (-\al' s/2)p^2} \delta(p\cdot u_1)\delta(p\cdot u_2)\nn\\
&\sim & {s^{-(D-2)/2}\over\sinh\chi}
  e^{-(f_2-f_1)_\perp^2/(2\al' s)}\period\label{ampzero}
\eqaend
The exponent only contains $(f_2-f_1)_\perp$, the components transverse 
to the trajectory plane, and this agrees with the discussion in 
 section 2.1. The  factor $1/\sinh\chi$ comes from the 
 momentum integral in the trajectory plane, 
 $\int d^2p \delta(p\cdot u_1)\delta(p\cdot u_2)$, which gives 
the inverse of the area spanned by the non-orthogonal 
unit vectors $u_1$ and $u_2$.  The expression (\ref{ampzero}) agrees 
 completely with the corresponding zero mode contribution 
 computed by  the path integral method. \par
Next, we turn to  the non-zero modes. 
Propagation through the proper time $s$ gives a factor 
  $e^{-2ns}$ inside the sum in the exponent of the boundary ket and 
 hence what we need to compute is
\eqabegin
\calV_{nz}&=&   \bra{0}\exp\left( -\sum_{n\ge1} {1\over n} 
 \al^\mu_n D^\nu_\mu(\fdot_2) \altil_{n, \nu}\right) \nn\\
&& \qquad \cdot \exp\left(-\sum_{n\ge1} {1\over n} e^{-2ns}
 \al^\mu_{-n} D^\nu_\mu(\fdot_1) \altil_{-n, \nu}\right)\ket{0}\period
\eqaend
This can be evaluated by normal ordering of the oscillators. 
In  the Appendix, we derive a complete normal-ordering 
formula\footnote{For the present purpose, we only need the formula for 
 the vacuum expectation value, which is certainly well-known. 
 The full formula given below, however,  is needed  in more involved  
 computations (such as those of  Green's functions, etc.). Since, to our 
 knowledge, it has not been recorded in the literature, we provide a 
 derivation in the Appendix.} 
 by  extending the method developed in \cite{Corrigan74}. 
Let $( a_i,a^\dagger_i)$ and $( \atil_i,\atil^\dagger_i)$ be two 
 independet sets of oscillators satisfying the usual (anti-)commutation 
relations  $\left[ a_i, \adag_j\right]_\pm = 
 \left[ \atil_i, \atildag_j\right]_\pm = \delta_{ij}$,  $A$ and $B$ be 
 arbitrary matrices, and write $aA\atil \equiv 
 \sum_{i,j} a_i A_{ij} \atil_j$ etc.. Then we have  
\eqabegin
&&  \exp\left(a A  \tilde{a}\right) 
\exp\left( {\tilde{a}}^\dagger 
 B a^\dagger\right) \nn\\  
&& = \left[ \det(1-\ep BA)\right]^{-\ep} 
  \exp\left( {\tilde{a}}^\dagger 
 (1-\ep BA)^{-1} Ba^\dagger \right)
\exp\left( aA (1-\ep BA)\tilde{a}\right) \nn\\
&& \qquad \cdot   
 \exp\left(-a^\dagger\ln(1-\ep B^T A^T) a-
 {\tilde{a}}^\dagger \ln(1-\ep BA) \tilde{a}\right) \comma \label{noformula}
\eqaend
where $\ep =+1\, (-1)$ for commuting (anti-commuting) oscillators.
Setting 
\eqabegin
 a_{n,\mu} &=& {i \over \sqrt{n}} \al_{n,\mu}\comma 
\qquad a^\dagger_{n,\mu} = {1\over i\sqrt{n}} \al_{-n,\mu}\comma \\
A_{n\mu, \ell\nu} &=& D_{\mu\nu}(f_2)\delta_{n\ell}\comma \qquad 
B_{n\mu, \ell\nu} = e^{-2ns}D_{\mu\nu}(f_1)\delta_{n\ell}\comma 
\eqaend
we  readily obtain 
\eqabegin
\calV_{nz} &=& \prod_{n=1}^\infty 
\left[\det\left(  1-D_1D_2 e^{-2ns}\right)  \right]^{-1} \comma 
\label{anz1}
\eqaend
where  the determinant  here is  over the space-time indices only and 
 we have written $D_i$ for $D_\mu^\nu(\fdot_i)$. 
It is easy to show that in our orthonormal basis $\left\{ u_1, \util_2,
u_I\right\}$ defined in (\ref{stdbasis}), the $D_i$'s have the following 
 representation:
\eqabegin
 D_i &=& d_i  \oplus (-{\bf 1}_{D-2})\comma  \\
d_1 &=&  \matrixii{1}{0}{0}{-1}  \comma  \\
d_2 &=&  \matrixii{\cosh 2\chi}{\sinh 2\chi}{-\sinh 2\chi}{-\cosh 2\chi} 
 \comma 
\eqaend
where ${\bf 1}_{D-2}$ stands for the unit matrix in the $D-2$ 
 dimensional space spanned by $u_I$. 
Therefore 
\eqabegin
 D_1D_2 &=&  d_1d_2  \oplus {\bf 1}_{D-2} =
 \matrixii{\cosh 2\chi}{\sinh 2\chi}{\sinh 2\chi}{\cosh 2\chi} 
\oplus {\bf 1}_{D-2}\period
\eqaend
Note that the matrix $d_1d_2$ is of the form of a Lorentz boost  
 in two dimensions and hence its eigenvalues are $e^{\pm 2\chi}$.
Therefore we get 
\eqabegin
\calV_{nz} &=&   \prod_{n=1}^\infty   \left[\det\left(  1-D_1D_2 e^{-2ns} 
\right)  \right]^{-1} \nn\\
&=& \prod_{n=1}^\infty  \left(1-e^{-2ns+2\chi}\right)
 \left(1-e^{-2ns-2\chi}\right) (1-e^{-2ns})^{-(D-2)} \period
\eqaend
\par
Finally, consider the ghost contribution. As is well-known, due to the 
 existance on the cylinder of one Teichm\"uller parameter and 
 one conformal symmetry (rotation), we need to insert a zero mode
 $U_0=(c_0-\tilde{c}_0)(b_0+\tilde{b})$  in order to get a non-vanishing 
 inner product\cite{CallanII}. 
Then,  applying (\ref{noformula}) with $\ep=-1$, we get 
the familiar answer 
\eqabegin
\calV_{gh} &=& \bra{B_{gh}}e^{-s(L_0^{gh}+\tilde{L}_0^{gh})} U_0
 \ket{B_{gh}} =\prod_{n=1}^\infty (1-e^{-2ns})^2\period
\eqaend
\par
Assembling altogether we thus obtain
\eqabegin
\calV_0(f_1, f_2) &\sim & \int_0^\infty ds e^{2s} 
 s^{-(D-2)/2} e^{-(f_1(t_1)-f_2(t_2))_\perp^2/2\al' s} \nn\\
&& \quad \cdot {1\over \sinh\chi}  \prod_{n=1}^\infty 
 (1-e^{-2ns})^{-(D-4)}
 \left(1-e^{-2ns+2\chi}\right)^{-1}
 \left(1-e^{-2ns-2\chi}\right)^{-1} \comma 
\eqaend
which, for $D=26$, is identical to the lowest order result (\ref{loamp}) 
obtained by the path integral method. 
\par
We can also compute the Green's function $G_{ij}(z,z') $ (see 
 (\ref{2dgrfn})),  satisfying the  rather complicated boundary conditions,  
 by using the boundary state representation.  Since the one we need  
 is in the space of non-zero modes,  the relevant quantity is 
 the non-zero mode part of the amplitude with two closed string 
 tachyon vertices inserted
\eqabegin
 \calA_{nz}(k, k') &=& \langle B_{nz};f_2| 
 e^{ik\cdot X^{(+)}(z)} e^{ik\cdot X^{(-)}(z)}
e^{ik'\cdot X^{(+)}(z')} e^{ik'\cdot X^{(-)}(z')}|B_{nz};f_1\rangle\comma 
\eqaend
where $X^{(\pm)}$ are the non-zero mode parts of the closed string 
 cooridnate given by $X^{(\pm)}(z) =\mp i\sum_{n=1}^\infty
 (1/n)(\al_{\mp n}z^{\pm n} + \altil_{\mp n}\bar{z}^{\pm n})$. 
To compute this amplitude, the normal-ordering formula (\ref{noformula})
 is indispensable. As we wish to check only the non-trivial components 
 $G_{ij}(z,z')$ with $i,j$ in the trajectory plane, we restrict the momenta 
 $k$ and $k'$ to have   components  only in this plane. Then, 
combining with appropriate use of coherent 
 state method, we obtain, after a long calculation,  
\eqabegin
\calA_{nz}(k,k') &=& D \exp\left( 4 k^i G_{ij}(z,z') {k'}^j 
\right) \comma 
\eqaend
where $D$ is the determinant factor (\ref{anz1}), 
  $k^i$ stands for $ k\cdot u^i$ etc., and   
 $G_{ij}(z,z')$  is precisely the one that we computed in the previous 
 section by the path integral method.  
\section{ Operator Formalism in Open String Channel}
\setcounter{equation}{0}
\subsection{Amplitude in the open string channel}
In the previous two sections, we have computed the  D0-D0 scattering 
 amplitude essentially from the closed string channel. An important 
 question is to see how, in the low energy domain, the result 
 can be understood  in terms of a spontaneously broken 
  effective gauge theory\cite{Witten9510}, 
which is based on the open channel picture. 
Although there is no doubt that this gauge theory description is
 basically correct and is extremely useful,  how it should work for 
  D-particles with finite mass (and hence with recoil) is 
 still a non-trivial issue.  As was already emphasized in the introduction, 
it must be intimately related with the \lq\lq non-commutative nature
 of spacetime" and clarifying  this  connection would help understand 
 the physical meaning of this intriguing concept. This makes the 
 following  re-derivation of the amplitude in the open channel  a  
significant exercise. 
\par
 Before taking up this task, 
 let us first perform the modular transformation of the lowest order 
 amplitude (\ref{loampth}) to see what to expect. 
Introducing the usual open channel modular parameter $w$ as 
\eqabegin
w&\equiv & e^{2\pi^2/s} \comma \qquad 
{-\ln q\over \pi} = {-2\pi \over \ln w}\comma 
\eqaend
and using the standard formulas for the modular transformation 
 of the $\theta$-functions, 
the expression  (\ref{loampth})   for $D=26$ is transformed into  
\newpage
\eqabegin
\calV_0(f_1,f_2) &\sim&  \int_0^1 {dw\over w^2} 
w^{(f_1-f_2)_\perp^2/(4\pi^2 \al')} w^{\nu(1-\nu)/2}  f(w)^{-24}\nn\\
&& \times {1 \over 1-w^{\nu}}\left[ {f(w)^2 \over \prod_{m=1}^\infty 
(1-w^{m+\nu })(1-w^{m-\nu})}\right] \period \label{opamp}
\eqaend
where 
\eqabegin
\nu &\equiv & -i {\chi \over  \pi } \comma \qquad 
 f(w) \equiv \prod_{m=1}^\infty (1-w^m) \period \label{nudef}
\eqaend
This indicates that apparently the open string spectrum is a peculiar one
 in that (i) the Casimir energy $\nu(1-\nu)/2$ is complex, (ii) two types of
 non-zero modes exist with complex energy levels  $m+\nu$ and
  $m-\nu$,
 and (iii)  an additional  excitation appears  with pure 
 imaginary energy $\nu$.  By performing 
 the quantization in the open string channel, 
we should be able to understand  the meaning of these modes 
 and reproduce  the amplitude in the form of the partition function 
\eqabegin
\calZ_{open} &= &  -\half \trace\ln (L_0-1) =\half 
 {dw \over w}\left( {-1\over \ln w}\right) \trace\, w^{L_0-1}\period 
\label{zopen}
\eqaend
%
\subsection{Quantization, spectrum and the amplitude}
Let us now perform the quantization in the open string channel
\footnote{The calculation will turn out to be 
  quite similar to the one perfomed 
 in \cite{Bachas}  (see also \cite{CallanII,Nesterenko}) 
but we will discuss the subtleties involved 
 and their physical interpretation in much more detail.}. 
We will employ the usual  strip coordinate $(\sigma, \tau)$, 
 where $0\le \sigma\le \pi$. As in the calculation in the closed  
 channel, we split the string coordinate into the classical   
 part $X_{cl}^\mu$ and the quantum fluctuation part $\xi^\mu$. 
In the present coordinate system, the classical solution takes the 
 form 
$ X^\mu_{cl} (\sigma, \tau) = {\sigma \over \pi}(f^\mu_2(t_2)
 -f^\mu_1(t_1)) 
 + f^\mu_1(t_1)$, 
which describes a string stretched from $f_1^\mu (t_1)$ to $f_2^\mu(t_2)$.
Remembering that the components of $f_2-f_1$ in the trajectory 
 plane must vanish by consistency,  
the energy associated with this stretching is 
$ H_{cl}= (f_1(t_1)-f_2(t_2))_\perp^2/(4\pi^2 \al')$, 
which   immediately gives the factor $w^{(f_1-f_2)_\perp^2/(4\pi^2\al')}$ 
appearing  in (\ref{opamp}). \par
As for the fluctuation part, we  will continue to use the orthonormal frame 
 defined in (\ref{stdbasis}).  
 Since the $D-2$ components $\xi_I$ transverse 
 to the trajectories satisfy the usual 
 Dirichlet conditions and their quantization is standard, we will concentrate
 on the tangent components $\xi_1$ and $\xi_2$. By now the appropriate 
 boundary conditions shoud be familiar:  
\eqabegin
&& \del_\sigma \xi_1(0, \tau) = 0 \comma \quad 
\xi_2(0,\tau) = 0 \comma   \\
&& \xi_1(\pi, \tau) \sinh \chi +\xi_2(\pi, \tau) \cosh \chi =0 \comma \\
&& \del_\sigma \xi_1(\pi, \tau) \cosh \chi +\del_\sigma
 \xi_2(\pi, \tau) \sinh \chi=0\period  \\
\eqaend
It is not difficult to find the solutions of the free wave equation  
 satisfying  these conditions and expand $\xi_1$ and $\xi_2$ in terms
 of these modes. A convenient way of writing the expansions is 
\eqabegin
\xi_1 &=&{\sqrt{\al'}\over \lam^+_0}  \left( e^{(\chi/\pi) \tau}\bebar
 + e^{-(\chi/\pi) \tau}\be\right)
 \cosh {\chi \over \pi}\sigma \nn\\
 && + \sqrt{\al'} \sum_{n\ge 1, \ep=\pm} 
 {1\over \lam^\ep_n}
 \left( \al^\ep_n e^{-i\lam_n^\ep \tau} 
- \al^\ep_{-n} e^{i\lam_n^\ep \tau} \right) 
 \cos \lam^\ep_n \sigma \label{xione}\\
\xi_2 &=& -{\sqrt{\al'}\over \lam^+_0}
\left( e^{(\chi/\pi) \tau} \bebar + e^{-(\chi/\pi) \tau}\be\right)
 \sinh {\chi \over \pi}\sigma \\
 && +i \sqrt{\al'} \sum_{n\ge 1, \ep=\pm}
 {\ep\over \lam^\ep_n} \left( \al^\ep_n e^{-i\lam_n^\ep \tau} 
-\al^\ep_{-n} e^{i\lam_n^\ep \tau} \right) 
 \sin \lam^\ep_n \sigma \label{xitwo}
\eqaend
where $\lam^\pm_m$ are given by 
\eqabegin
\lam^\pm_m &=& m\pm i{\chi \over \pi}\ \period
\eqaend
As was already anticipated  from  the modular transformed form
 (\ref{opamp}),  
 the eigenfrequencies are complex and the modes grow or diminish as 
 functions of $\tau$. Technically, this comes about because the 
 boundary conditions involve hyperbolic functions of $\chi$ while 
 the oscillatory part is trigonometric. Physical origin  will be 
 discussed  after  we finish the quantization. \par
Since the creation and the annihiliation operators should have opposite 
 $\tau$-dependence, one expects that the conjugate pairs are 
$(\al^+_n, \al^+_{-n})$, $(\al^-_n,\al^-_{-n})$, and $(\be, \bebar)$. 
Indeed it is easy to check that the canonical quantization conditions 
\eqabegin
 \left[\dot{\xi_i}(\tau,\sigma), 
 \xi_j (\tau,\sigma')\right]&=&{2\pi \al'\over i} \delta(\sigma -\sigma')
\eta_{ij}  
\eqaend
are satisfied if we impose the following commutation relations:
\eqabegin
\com{\al^+_m}{\al^+_{-n}}&=& \lam^+_m \delta_{mn}\comma \quad 
\com{\al^-_m}{\al^-_{-n}}= \lam^-_m \delta_{mn}\comma \quad 
\com{\be}{\bebar}= \lam^+_0\period
\eqaend
It is then a bit tedious but straightforward to compute the Virasoro 
 operators $L_n^{12}$ for the $\xi_1$-$\xi_2$ system by the expansion
\eqabegin
\sum_n L_n^{12} e^{-in(\tau+\sigma)}&=& {1\over 4\al'}\left[
 -(\del_\tau\xi_1+\del_\sigma\xi_1)^2 +(\del_\tau\xi_2+\del_\sigma\xi_2)^2
\right]\period
\eqaend
(One can check that on the right hand side  terms which do not have 
 the dependence $e^{-in(\tau+\sigma)}$ all cancel.) The result is,
  with $n\ge 1$, 
\eqabegin
L_0^{12} &=& -\bebar \be 
 + \sum_{m\ge 1} (\al^+_{-m}\al^+_m +\al^-_{-m}\al^-_m )
 +\half\nu(1-\nu) \label{lzero12} \comma \\
L_n^{12} &=&\bebar \al^-_n -\be\al^+_n
 + \sum_{l+m=n; l,m\ge 1}\al^+_l\al^-_m  \nn\\
&&  + \sum_{m\ge 1}
 (\al^+_{-m} \al^+_{m+n} +\al^-_{-m} \al^-_{m+n} ) \comma \\
L_{-n}^{12} &=&  \bebar\al^+_{-n} 
-\be\al^-_{-n}
 + \sum_{l+m=n; l,m\ge 1}\al^+_{-l}\al^-_{-m} \nn\\
&&  + \sum_{m\ge 1, m-n\ge 1}
 (\al^+_{-m} \al^+_{m-n} +\al^-_{-m} \al^-_{m-n} )\period
\eqaend
The shift $\nu(1-\nu)/2$ in  $L_0^{12}$, which takes exactly the form expected, 
  has been determined so that these operators
 satisfy the usual form of the Virasoro algebra (with central charge equal to
 $2$):
\eqabegin
\com{L_m^{12}}{L_n^{12}} &=& (m-n)L_{m+n}^{12} + \delta_{m+n,0}
{2\over 12}(m^3-m)\period
\eqaend
Also note that $L_0^{12}$ is hermitian  with this shift included. 
\par
Now,  since $L_0^{12}$ appears to be diagonal in the number operators 
 for the oscillator modes, one may expect  that the trace over the 
 modes,  $\trace w^{L_0^{12}}$,  readily
 leads to the expressions in (\ref{opamp}). 
This naive reasoning must however be carefully examined  because of 
 the unusual hermiticity properties of the oscillators. From the explicit form
 of $\xi_1$ and $\xi_2$ given in (\ref{xione}) and (\ref{xitwo}),  
 one  finds that the reality of these fields requires   
\eqabegin
 (\al^+_m)^\dagger &=& -\al^-_{-m}\comma \qquad 
(\al^-_m)^\dagger = -\al^+_{-m} \comma \\
\be^\dagger &=& \be\comma \qquad \bebar^\dagger = \bebar\period
\eqaend
This  shows that, while $\beta$  and $\bebar$ are like a coordinate
 and its conjugate momentum, $\al^\pm_n$ are interchanged under hermitian 
 conjugation  and hence cannot be identified as  usual oscillators. \par
To study the structure of the Hilbert space for these unusual
 $\al^\pm_n$ oscillators, let us concentrate on  a particular 
 level $n$ and 
consider the  system defined by 
\eqabegin
\com{a}{\atil} &=& \lam 
\comma\qquad  \com{b}{\btil} =\lam^\ast \comma  \\
a^\dagger &=&- \btil\comma \qquad  \atil^\dagger =-b \comma \\
H &=& \atil a +\btil b\comma 
\eqaend
where $\lam $ is a non-vanishing complex number. 
  The Hamiltonian $H$ is 
 obviously hermitian and the system is completely consistent. Suppose we  
 define the vacuum state $\psi_{0,0}$ by 
\eqabegin
 a\psi_{0,0}&=& 0\comma \qquad b\psi_{0,0}=0\period
\eqaend
Then we can build a general excited state $\psi_{m,n}$ by 
\eqabegin
 \psi_{m,n}&=& N_{m,n}\atil^m \btil^n \psi_{0,0} \comma 
\eqaend
where $N_{m,n}$ is a normalization constant.
 This is clearly an eigenstate of $H$ with 
 the eigenvalue $m\lambda +n\lambda^\ast$, which is {\it complex}  
 unless $m=n$. It is well known that a hermitian operator can have complex
 eigenvalues if and only if the norm of the corresponding eigenstates vanish
\cite{Arisueetal}. Indeed we have 
\eqabegin
 (\psi_{m,n}, \atil a \psi_{m,n})&=& m\lam (\psi_{m,n}, \psi_{m,n})\nn\\
 &=& (\btil b \psi_{m,n}, \psi_{m,n})=n\lam (\psi_{m,n}, \psi_{m,n})\comma 
\eqaend
and the norm $(\psi_{m,n}, \psi_{m,n})$  vanishes for $m\ne n$. By a similar
 manipulation, one can easily show that the state which has non-vanishing 
 inner product with $\psi_{m,n}$ is $\psi_{n,m}$. In other words, the metric 
  of this Hilbert space is non-diagonal\footnote{This means  that 
 the Hilbert space has negative norm states, but this is simply due 
 to the timelike nature of  $\xi_1$.}. If we choose the normalization 
 so that $(\psi_{n,m}, \psi_{m,n})=1$, then the trace of an operator $\calO$
 should be defined as 
\eqabegin
\trace\, \calO &\equiv & \sum_{m\ge 0, n\ge 0}(\psi_{n,m}, \calO \psi_{m,n})
\period
\eqaend
Hence the partition function is computed as 
\eqabegin
\trace\, w^H &=& \sum_{m\ge 0, n\ge 0}(\psi_{n,m}, w^H \psi_{m,n})\nn\\
&=& \sum_{m\ge 0, n\ge 0} w^{m\lam +n\lam^\ast}=\left(\sum_{m=0}^\infty
 w^{m\lam}\right)\left(\sum_{n=0}^\infty w^{m\lam^\ast}\right)\nn\\
&=& (1-w^\lam)^{-1}(1-w^{\lam^\ast})^{-1}\comma  \label{tracewh}
\eqaend
which turned  out to be identical  with the result expected by the naive
 reasoning. \par
To understand the physical meaning of this structure as well as  to check 
 the result in a different way, it is instructive to form new oscillators 
 which satisfy the conventional hermiticity property. For 
 example, if we define $A,A^\dagger, B,B^\dagger$ as 
\eqabegin
 A &=& {1\over \sqrt{2l}|\lam|} (\lam a + \lam b) \\
A^\dagger &=&{-1\over \sqrt{2l}|\lam|}(\lam^\ast \atil + \lam^\ast
\btil) \\
 B &=& {1\over \sqrt{2l}|\lam|} (\lam^\ast a - \lam b) \\
B^\dagger &=&{1\over \sqrt{2l}|\lam|}(\lam^\ast \atil - \lam\btil)\comma  
\eqaend
where $l $ is the real part of $\lam$ and is taken positive,  
 they satisfy the standard  commutation relations   
$ \com{A}{A^\dagger} =-1 \comma 
 \com{B}{B^\dagger} = 1 $. 
In terms of these oscillators, the Hamiltonian becomes  
\eqabegin
H &=& \atil a + \btil b \nn\\
  &=& {|\lam|^2 \over l} \Biggl\{ B^\dagger B -\half \left(
 {\lam\over \lam^\ast}+{\lam^\ast \over \lam}\right)A^\dagger A \nn\\
 && +\half \left( 1-{\lam \over \lam^\ast}\right)A^\dagger B 
 +\half \left( 1-{\lam^\ast \over \lam}\right)B^\dagger A\Biggr\}\period 
\eqaend
Thus,  in this representation  the metric of the Hilbert space
 is diagonal  but instead the Hamiltonian is non-diagonal for  
 complex $\lam$.  As one can easily check, 
  $A$ and $B$ are, respectively,  the oscillators which describe the usual particle modes for $\xi_1$ and  $\xi_2$ when $\lam$ becomes real. 
This means that,  except for a special situation where  $\chi$ vanishes,  
  the usual particle-like modes are \lq\lq unstable" and as time goes on they 
 incessantly transform into each other. Nevertheless, since the Hamiltonian
 is hermitian, the total probability is conserved and the process is 
 unitary.  \par
As for the calculation of the trace, 
 although a bit tedious, one can reproduce the previous result  
(\ref{tracewh}) using the above non-diagonal Hamiltonian 
(for example by  the use of the coherent state method).
Finally, applying this to the original problem, we find that the contribution 
 of the $\al^\pm_m, (m\ne 0)$ oscillators to the trace $\trace\, w^{L_0^{12}}$ 
 is given by 
\eqabegin     
\prod_{m=1}^\infty 
(1-w^{m+\nu})^{-1}(1-w^{m-\nu})^{-1}\comma 
\eqaend
as anticipated. 
\par
Let us next consider the $\beta$-$\bebar$ system, which is characterized by 
\eqabegin
 \com{\be}{\bebar} &=& i{\chi\over \pi}  \comma \\
\be^\dagger &=& \be\comma \qquad \bebar^\dagger = \bebar\comma \\
H &=& -\bebar\be-{i\chi \over 2\pi}= H^\dagger   \period
\eqaend
Before performing  a proper computation of the trace $\trace\, w^H
 =\trace\, e^{-\pi t H}\comma t=-\ln w/\pi$ 
 for this system, let us  describe the essence of the physics by looking 
 at the small $\chi$ limit.  In this limit, the boundary conditions for 
 $\xi_1$ and $\xi_2$ become almost Neumann and Dirichlet 
 respectively. Indeed if we introduce  $q$ and $p$ by 
\eqabegin
 q &=& {\sqrt{\al'}\over  i}(\be +\bebar) \comma 
 \qquad p = {i\over 2\sqrt{\al'}}(\be -\bebar)\comma \\
 \com{p}{q} &=& i{\chi\over \pi} \comma \label{pqcom}
\eqaend
$\xi_i$ can be written in the form  
\eqabegin
 \xi_1 &=& {\pi q\over \chi} + 2\al' p\tau  + \calO(\chi^2) \comma \\
\xi_2 &=& -q \sigma + \calO(\chi^2)\comma 
\eqaend
with the energy   
\eqabegin
 H &=& {1\over 4\al'}q^2 -\al' p^2\period \label{behamil}
\eqaend
This corresponds to the following picture: When $\chi$ is small, 
 while  the $\sigma =0$ end of the string can fluctuate strictly along 
 the $u_1$  direction,  the  $\sigma=\pi$ end may slide along the direction 
 which is slightly tilted from  $u_1$ into the perpendicular 
 $\tilde{u}_2$ direction.   Thus, the string, as a whole without 
 oscillation, can move into the $u_1$ direction almost freely but this 
 motion is accompanied by  a  stretching  or shrinking  in the $\tilde{u}_2$  direction.  This costs energy and results in the restoring potential seen in  the Hamiltonian. Now from (\ref{pqcom}) we see that as $\chi $ 
 tends to vanish,  $p$ and $q$ will become independent. 
Therefore, in this limit we get 
\eqabegin
\trace\, e^{-\pi tH} &\longrightarrow &  {\pi\over \chi} 
 \int dq  dp  e^{-\pi tq^2/4\al'} 
 e^{\pi t \al' p^2} 
 \sim {1 \over \chi  \ln w} \comma 
\eqaend
where the  factor  in front of the integral compensates
 the normalization of the commutator  (\ref{pqcom}).  
One can easily check that this agrees    with 
 the $\chi \rightarrow 0$ limit
 of the expression $1 /(1-w^\nu)$ in (\ref{opamp}).  
\par
Let us now compute $\trace\, e^{-\pi tH}$ for general $\chi$. 
In order to build  a well-defined Hilbert space, we shall define a 
conventional  oscillator pair $(a,a^\dagger)$ by linear 
combinations of $\be$ and $\bebar$.
A simple choice that realizes $\com{a}{a^\dagger}=1$, $ (a)^\dagger =
 a^\dagger$ is   
\eqabegin
a &=& {1\over \sqrt{2\chi/\pi}} (\be +i\bebar) \comma \qquad 
a^\dagger = {1\over \sqrt{2\chi/\pi}} (\be -i\bebar)\period
\eqaend
(More general ones  are unitarily equivalent to this.)  Then the Hamiltonian
 takes the form 
\eqabegin
 H &=& 
{i\chi \over 2\pi}(a^2 -{a^\dagger}^2+1)\comma  \label{hamilaa}
\eqaend
which is not number-diagonal. The relevant trace $\trace\, e^{-\pi tH}$  
can nevertheless be computed in various ways. One way 
 is to first perform the normal ordering by a  technique similar to the one 
 explained in the Appendix, which gives 
\newpage
\eqabegin
 e^{-\pi tH} &=& \delta e^{\alpha {a^\dagger}^2} e^{\be a^2} e^{\gamma a^\dagger a} \comma \\
\where \delta &=& e^{-i\chi t /2} (\cos \chi t)^{-1/2}\comma  \\
\al &=& {i \over 2} \tan \chi t\comma  \\
\be &=& -{i \over 2} \sin\chi t \cos \chi t \comma \\
\gamma &=& -\ln (\cos\chi t)\period
\eqaend
Then the  trace of this expression is easily obtained by using the coherent 
state  method. The result is 
\eqabegin
 \trace\, e^{-\pi tH}&=& \delta \left[ (1-e^\gamma)^2 -4\al\be e^{2\gamma}
\right]^{-1/2} \\
&=& {e^{-i\chi t /2}\over 2i\sin{\chi t\over 2}} = {1\over 1-w^\nu}
\period 
\eqaend
Thus we obtain  the desired expression. \par
We  now briefly describe the contributions of  the $D-2$ transverse 
 components and of the ghosts. \par 
As for  $\xi_I$'s,  which satisfy the standard Dirichlet boundary conditions, 
 there are no zero modes and their contribution therefore is simply 
$ f(w)^{-(D-2)}$.  \par
Finally the ghosts. Since they are not affected by the background, 
 the analysis is the same as in the usual case, which was 
 described in detail  in the appendix of \cite{CallanII}. 
The essential point is the following. Recall that due to the presence 
 on a cylinder of a Teichm\"{u}ller parameter and a conformal symmetry
the calculation using the boundary states in the closed string channel 
 required an insertion of appropriate  zero modes, which 
 is equivalent to an insertion of the  ghost number operator. 
 When converted  into the open string channel, the effect of this 
 insertion produces a factor of $\ln w$.  In this way, one obtains 
 the formula connecting the ghost partition function in the closed 
 channel and that in the open channel:
\eqabegin
 q^{1/6} f(q^2)^2 &=& w^{1/12} f(w)^2 (-\ln w /2\pi)\period 
\eqaend
This, however,  is  nothing but the famous modular transformation formula 
for the (square of ) the Dedekind $\eta$ function.  Since the powers 
 of $q$ and $w$ appearing in this formula 
  have already been taken care of by the  
 total intercept of the Virasoro operator,
  the contribution we must add is $f(w)^2(-\ln w/2\pi)$. 
 In particular, this factor of $\ln w$ is exactly what is needed to 
 cancel the $1/\ln w$ in (\ref{zopen}). \par
Thus, putting everything together, we have reproduced  
all the factors that appear in the scattering amplitude as seen in 
 the open channel. 
  The important lesson we learned from this 
 exercise is that  non-trivial mixing of particle modes occur in this
 channel.  We are intending to  understand this phenomenon 
 from the point of view of effective gauge theory in a future report.         
\section{Discussions}
\setcounter{equation}{0}
In this article, we have computed the amplitude for the scattering 
 of two D-particles in bosonic string theory, 
 where D-particles themselves are quantized. 
 We employed  three different 
 methods, namely, the path integral, the boundary state, and the 
 operator formalism in the open string channel, and cross-checked 
 the result.  
 The emphasis was on the path integral method, which is 
 conceptually most complete in formulating the problem. Especially, 
 it is only with this method that we can clearly grasp the nature of  
the  
 approximation used and compute the corrections  in a systematic manner. 
As far as the computation in the lowest order 
(in the acceleration $\ddot{f}_i $) is concerned, the use of the 
 boundary state representation of the interaction vertex  appears to 
 be most efficient.  On the other hand, the operator method in 
 the open string channel  reveals the occurence of 
  incessant transitions among the 
 excitations of the open string that  connect the D-particles and 
 perhaps a  
 deeper understanding  of this behavior will  be important 
 in unravelling the connection with the description in terms of 
 the spontaneously broken gauge theory.  \par
In any event, the fact that the interaction between 
 {\it quantum} D-particles through all the excitation modes of 
 a string is fully consistent is rather remarkable. 
This is because in the present approach 
D-particles are no longer backgrounds but are new independent 
 entities which can coexist with strings. This raises a puzzling 
 question\footnote{This question was raised 
 by N.~Ishibashi during our discussion.}: Is it not true that 
  they are supposed to be solitons of string theory 
 and hence they should  not represent new degrees of freedom ? 
 This may indicate  that 
 some important non-perturbative  consistency condition 
linking the two is still  missing. On the other hand, 
 the conjecture\cite{Witten953} that the D-particles are the Kaluza-Klein 
 modes of 11 dimensional M theory indicates that they could indeed be 
degrees of freedom independent of those of strings. The answer is 
 not yet known, but it is certainly an important question in the 
 light of consistency of our result. \par
In this article, we have presented the calculation only for bosonic 
 string theory and have left out the more important case of 
 superstring. The reason is that in the latter case, a satisfactory 
 analysis requires the understanding of not only the scattering 
 of bosonic D-particles but also that involving fermionic 
 superpartners as well and this appears to be non-trivial. 
In the path integral approach, we need to  consider supertrajectories 
 and their quantization and this is known to be  difficult. 
This matter is under investigation and we hope to be able to 
 report our progress in the near future. \par
Nevertheless, if one is content 
 with the amplitude for  two bosonic quantum D-particles scattering 
into again two bosonic ones, we can present a well-educated guess. 
This is due to the fact that 
 in the bosonic string case (at least for the lowest order 
 approximation) there emerged  a 
 simple rule to go from the forward scattering 
 amplitude for infinitely heavy D-particles to our general amplitude 
 for quantized D-particles with finite mass and it is almost 
 obvious that this rule should continue to hold for the superstring case. Assuming this to be the case, all we have to do is to write down 
 the result obtained in \cite{Bachas},  make one simple 
 substitution and a Fourier transformation, and  take 
into account the quantum indistinguishability discussed 
in subsection 2.4.  Explicitly, 
the general on-shell amplitude in the open channel representation 
 should read 
\eqabegin
\calA_{super}(p_i,p'_i) &=& 
\calA_{super}(p_1,p_2; p'_1,p'_2) + \calA_{super}
(p_1,p_2; p'_2,p'_1) \comma \\
\eqaend
where 
\eqabegin
\calA_{super}(p_1,p_2; p'_1,p'_2) &\sim & 
\int_0^\infty {dt\over t} t^{-4} e^{-\al'\pi (p'_1-p_1)^2/t} 
 {\theta_1'\left(0|{it\over 2}\right) \over 
\theta_1\left({\chi t\over 2\pi}|{it\over 2}\right)} \nn\\
&& \quad \times \sum_{\al=2,3,4} \theta_\al
\left({\chi t\over 2\pi}|{it\over 2}\right)
\theta_\al^3
\left(0|{it\over 2}\right) \eta^{-12}\left( {it \over 2}\right) \period
\eqaend
In this formula, $\chi$ is as given in (\ref{coshchi}) $\sim$  (\ref{ui}) 
 and $e_2 = -e_3= e_4 =-1$. 
We intend to discuss in detail how this general formula fits 
 with the super-Yang-Mills description in a future communication. 
\par
We  finish by making an important observation  that the type of 
 amplitude for the basic process considered in this paper, 
 no matter how accurately computed in superstring theory and no matter 
 how small the string coupling constant is,   
 cannot be relied upon if one wishes to know the behavior of 
 the D-particles at very high energy. The difficulty stems from the 
fact that the annulus amplitude does not explicitly depend on the string 
 coupling constant $g_s$. Thus, one can insert \lq\lq annulus 
 interactions" anywhere one likes and produce 
  infinitely many diagrams, 
 such as the ones dipicted in Fig.4, which contribute at the 
 same order in $g_s$.
\begin{center} 
\epsfxsize=8cm
\quad\epsfbox{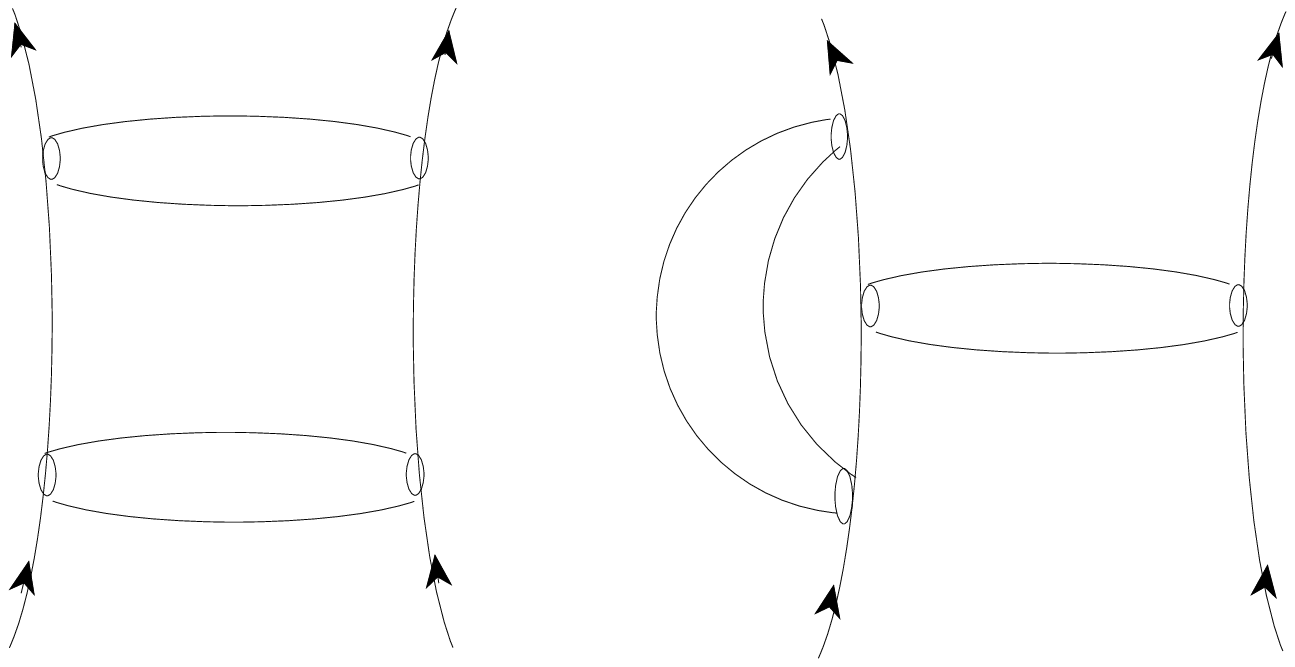} \\
{\small {\bf Fig.4} \quad Example of diagrams at the same order 
 in the string coupling as the basic one.}
\end{center}
\parn
%
%
%
 At energy much smaller than the 
 D-particle mass, the diagrams with extra 
 corrections are expected to be suppressed 
 by  powers  of $E^2 / m^2$ due to the presence 
 of internal propagators of D-particles. But as the energy becomes 
 comparable to $m$  all these diagrams would contribute equally 
 to the amplitude and computation appears to become intractable. 
 It would however  be extremely interesting if one can devise a method 
 to extract the leading high energy behavior out of these 
 diagrams just as the similar study  for ordinary strings
\cite{GrossMende} stimulated a lot of thinking about the 
 degrees of freedom of string theory at short distance. 
\par\bigskip\noindent
{\large\bf Acknowledgment}\par\smallskip\noindent
I am grateful to S.~Hirano for posing 
valuable questions and for sharing some thinking with me
 about the boundary state method, which
  helped sharpened my understanding of the results presented in this 
 paper.  
I also would like to thank my colleagues  M.~Kato and T.~Yoneya  
 for stimulating discussions on the present work and 
 on  the recent developments of string theory in general. 
A preliminary version of this work was presented in a workshop 
 \lq\lq Non-perturbative Methods in String Theory" held at KEK 
 in March, 1997. 
I  wish to thank the participants for useful comments.  
 Especially  conversations with  N.~Ishibashi  and H.~Kawai are 
 gratefully acknowledged. 
\setcounter{equation}{0}
\renewcommand{\theequation}{A.\arabic{equation}}
\par\bigskip\noindent
{\Large\bf Appendix:} {\large\bf \quad Normal Ordering Formula}
\parmedskipn
Let $( a_i,a^\dagger_i)$ and $( \atil_i,\atil^\dagger_i)$ be two 
 independet sets of oscillators satisfying the usual (anti-)commutation 
relations  $\left[ a_i, \adag_j\right]_\pm = 
 \left[ \atil_i, \atildag_j\right]_\pm = \delta_{ij}$. Throughout, upper
 (lower) sign refers to the anti-commuting (commuting) case. 
Consider the bilinears 
\eqabegin
 \al &=& aA\atil  \comma \qquad 
\be =\atildag B\adag  \comma 
\eqaend
where $aA\atil \equiv \sum_{i,j}a_i A_{ij} \atil_j$,  etc..  By extending  
 the idea of \cite{Corrigan74}, we will derive a formula which 
 gives the fully normal ordered form of $e^\al e^\be$. \par
Let $\lambda$ be a parameter and we seek the result in the following 
 form:
\eqabegin
 e^{\lam\al} e^{\lam \be} &=& \delta(\lam) e^{x(\lam)}
 e^{y(\lam)}e^{z(\lam)} \comma \\
 x(\lam) &=& \atildag X(\lam) \adag\comma  \\
y(\lam) &=&  aY(\lam) \atil \comma \\
z(\lam) &=& \atildag Z(\lam) \atil+\adag W(\lam) a  \period
\eqaend
Here $X(\lam), Y(\lam), Z(\lam), W(\lam)$ are matrices and $\delta(\lam)$
 is a prefactor, to be determined. Let us  define  
\eqabegin
\chi_1(\lam) &=& \delta(\lam) e^{x(\lam)} e^{y(\lam)} \comma \\
\chi_2(\lam) &=&e^{\lam\al} e^{\lam \be}e^{-z(\lam)}\period 
\eqaend
Then what we want to obtain is $\chi_1(\lam) =\chi_2(\lam)$ and  
this is achieved uniquely if we can satisfy 
\eqabegin
&& (i)\quad \chi_1(0)  = \chi_2(0) \comma \\
&&(ii)\quad \chi_1^{-1}(\lam) \chi'_1(\lam) =\chi_2^{-1}(\lam) \chi'_2(\lam) 
\comma 
\eqaend
where $\chi'_1 \equiv d\chi_1/d\lam$ etc.. Writing out the condition 
  $(ii)$ explicitly, we get 
\eqabegin
  {\delta'\over \delta} +
 \chi_1^{-1}  x'(\lam) \chi_1 +y'(\lam) 
&=& \chi_2^{-1}\al \chi_2 
+\chi_2^{-1}e^{\lam\al} \be e^{-\lam\al} \chi_2 -z'(\lam)
\comma \label{chitwo}
\eqaend
where  we have made an assumption 
 $\com{z(\lam)}{z'(\lam)} =0$. Its validity will be aposteriori justified.  Evaluation of the quantities appearing in (\ref{chitwo}) 
 is a bit tedious but straightforward. Then, by
equating the coefficients of the same operator structure, we get 
 the following 5 differential equations:
\eqabegin
&&(a)\quad  {\delta' \over \delta} = \lam \Tr (BA) 
+  \Tr(X'Y) \comma \\
&&(b)\quad  X' = e^Z \left( 1\mp \lam^2 BA\right)B e^{W^T} \comma \\
&& (c)\quad Y' \mp YX'Y = e^{-W^T} A e^{-Z} \comma \\
&& (d) \quad X'Y = \mp Z' -\lam e^Z BA e^{-Z}\comma  \\
&& (e) \quad YX' = {W'}^T -\lam e^{-W^T} AB e^{W^T}\period
\eqaend
From the  structure of these equations,  
 it is consistent to assume that 
 $X'Y, Z, X'A, BY$ are functions only of $BA$, while  
 $YX'$ and  $W^T$ depend only on  $AB$.
Let us take the trace of (d) and (e). We get 
\eqabegin
 \Tr(X'Y) &=& \Tr Z' -\lam\Tr(BA) \nn\\
 &=& \Tr W' -\lam \Tr (BA)  \period \label{trxy}
\eqaend
Thus we must have $\Tr Z' = \Tr W'$. Substituting (\ref{trxy})
 into (a) and integrating, we get 
\eqabegin
 \delta &=& e^{\Tr Z} = e^{\Tr W} \period \label{eqdelta}
\eqaend
This result, together with the previously mentioned dependence 
 on $AB$ and $BA$, prompts us to make a postulate   
\eqabegin
 Z &=& f(BA) \comma \qquad W^T = f(AB)\comma  \label{zwpost}
\eqaend
where $f(x)$ is some  power-expandable function. Then since
 $B f(AB) = f(BA)B$ holds, we have 
\eqabegin
 BW^T &=& ZB \comma \qquad W^T A= A Z \period
\eqaend
Using these postulates, the equations (b) $\sim$ (d) simplify to 
\eqabegin
&&(b)\quad  X' = e^{2Z} \left( 1\mp \lam^2 \xi\right)B \comma  \\
&& (c)\quad Y' \mp YX'Y = A e^{-2Z} \comma \\
&& (d) \quad X'Y = \mp Z' -\lam \xi \comma \\
&& \mbox{where}\qquad  \xi \equiv  BA\period
\eqaend
To solve them, first 
multiply (b) from right by $Y$. We get 
\eqabegin
 X'Y &=& e^{2Z} \left( 1\mp \lam^2 \xi\right) \eta \comma \\
\where \eta &\equiv & BY \period 
\eqaend
 Equating this with the RHS of  (d), we can solve for $\eta$ in terms of $\xi$:
\eqabegin
 \eta &=& (1\mp\lam^2\xi)^{-1} e^{-2Z} (\mp Z' -\lam \xi)\period
\label{eqeta}
\eqaend
Next multiply (c) from left by $B$ and substitute (d) for 
$X'Y$. One then obtains 
\eqabegin
 \eta' \mp \eta (\mp Z' -\lam \xi) &=& \xi e^{-2Z}\period
\eqaend
We may now eliminate $\eta$ by substituting (\ref{eqeta}). After 
 some calculation we get a differential equation for $Z$ of the 
 form 
\eqabegin
 (1\mp\lam^2\xi)({Z'}^2 -Z'') &=& \pm 2\lam\xi Z' \pm 2\xi \period
\label{difeqz}
\eqaend
The solution of this equation satisfying the proper boundary condition
 $Z(0)=0$ is given by 
\eqabegin
 Z &=& -\ln(1\pm \lam^2 \xi)\period
\eqaend
We then get from  (\ref{eqdelta})  and the postulate (\ref{zwpost})
\eqabegin
 \delta &=& \exp\left( \pm \half \Tr \ln (1\pm \lam^2\xi) \right) 
= \left[ \det (1\pm \lam^2\xi)\right]^{\pm 1/2}\comma \\
 W &=& f(B^T A^T) = -\ln\left( 1\pm \lam^2 B^T A^T\right)\period
\eqaend
The matrix $X$, satisfying $X(0)=0$, 
  is obtained by integrating (b). We get
\eqabegin
 X &=& \int_0^\lam du {1 \mp u^2 \xi \over (1\pm u^2 \xi)^2} B=
 {\lam \over 1\pm \lam^2\xi} B \period
\eqaend
To get $Y$, we recall $Y = B^{-1} \eta$, where $\eta$ is given 
 by (\ref{eqeta}). One immediately gets
\eqabegin
 Y &=& \lam A (1\pm \lam^2\xi) \period
\eqaend
The remaining equation (e), which now reads 
\eqabegin
 YX'&=& \mp {W'}^T -\lam AB \comma 
\eqaend
is automatically satisfied. Also one can easily check the validity 
 of the assumption made earlier, namely $\com{z(\lam)}{z'(\lam)} =0$. 
Since the solution of the system of equations with appropriate boundary conditions is  unique, we can justify all the postulates made above. 
\par
Putting all together, we get the normal-ordering formula quoted
 in the text:
\eqabegin
&&e^{aA\atil} e^{\atildag B \adag} \nn\\
&& \quad = \left[ \det (1\pm  BA)\right]^{\pm 1} 
 e^{ \atildag (1\pm BA)^{-1}B\adag} \nn\\
&& \qquad \cdot e^{ aA(1\pm BA)\atil} 
 e^{-\atildag \ln(1\pm BA) \atil}e^{-\adag
 \ln(1\pm B^T A^T) a} \period
\eqaend
\par
For completeness, we exhibit the formula for the case of one 
 set of oscillators, which can be obtained in a similar manner:
\eqabegin
&& e^{ \half aAa} e^{\half \adag B\adag} \nn\\
&& \quad = \left[ \det (1\pm  BA)\right]^{\pm 1/2} 
 e^{\half \adag (1\pm BA)^{-1}B\adag} \nn\\
&& \qquad \cdot e^{\half aA(1\pm BA)a} 
 e^{-\adag \ln(1\pm BA) a} \period
\eqaend
In this case, the matrix $A$ and $B$ are symmetric (anti-symmetric) 
 if the oscillators are commuting (anti-commuting). 

\end{document}